\documentclass[conference]{IEEEtran}
\ifCLASSINFOpdf
\else
\fi
\usepackage{array}
\usepackage{longtable}

\usepackage[caption=false,font=footnotesize]{subfig}

\usepackage{tikz}
\def\checkmark{\tikz\fill[scale=0.4](0,.35) -- (.25,0) -- (1,.7) -- (.25,.15) -- cycle;}

\usepackage{multirow}
\usepackage{hhline}
\usepackage{xcolor,colortbl}
\definecolor{Gray}{gray}{0.89}
\usepackage{longtable}
\usepackage{supertabular}
\usepackage{rotating}
\usepackage{pdflscape}
\usepackage{afterpage}
\usepackage{float}

\usepackage{amsmath,amssymb}
\usepackage{array}
\def\mybar#1{
{\color{green}\rule{#1cm}{5pt}}}

\def\mybarA#1#2{
{\color{green}\rule{#1cm}{5pt}\color{red}\rule{#2cm}{5pt}}}

\usepackage{lscape}
\usepackage{tikz}
\usepackage{blindtext}

\begin{document}
%
\title{Recent Trends in Wearable Computing Research: A Systematic Review}

\author{\IEEEauthorblockN{Vicente J. P. Amorim, Ricardo A. O. Oliveira, Maur\'icio Jos\'e da Silva}
\IEEEauthorblockA{Computing Department (DECOM)\\
Federal University of Ouro Preto (UFOP)\\
Ouro Preto, Brazil\\
Email: vjpamorim@ufop.edu.br, rrabelo@gmail.com, badriciobq@gmail.com}}


%


\maketitle

\begin{abstract}
Wearable devices are a trending topic in both commercial and academic areas. Increasing demand for innovation has led to increased research and new products, addressing new challenges and creating profitable opportunities. However, despite a number of reviews and surveys on wearable computing, a study outlining how this area has recently evolved, which provides a broad and objective view of the main topics addressed by scientists, is lacking. The systematic review of literature presented in this paper investigates recent trends in wearable computing studies, taking into account a set of constraints applied to relevant studies over a window of ten years. The extracted articles were considered as a means to extract valuable information, creating a useful data set to represent the current status. Results of this study faithfully portray evolving interests in wearable devices. The analysis conducted here involving studies made over the past ten years allows evaluation of the areas, research focus, and technologies that are currently at the forefront of wearable device development. Conclusions presented in this review aim to assist scientists to better perceive recent demand trends and how wearable technology can further evolve. Finally, this study should assist in outlining the next steps in current and future development.
\end{abstract}


\IEEEpeerreviewmaketitle

\section{Introduction} \label{sec:introduction}
During recent years, the advent of new technologies has increased the demand for wearable devices \cite{gartner2018}. Wearables currently vary from simple biometric signal monitoring products to a complex set of solutions. These devices can integrate with other personal electronics, providing real-time data gathering, monitoring, and entertainment. Smartwatches, fitness bands, and AR/VR glasses are some examples of products frequently found in physical and online stores \cite{statista2018}. 

Although the wearable computing concept is not new \cite{Sutherland:1968:HTD:1476589.1476686}\cite{Mann:1996:SCS:232014.232021}, this area has recently received more attention, mainly because of hardware evolutions. Ongoing electronic components miniaturization has contributed to reducing devices' size and encapsulation, allowing the design and creation of new products, and embracing unexplored areas. Devices have become more wearable and pervasive, have been gaining increasing attention over the past years, and represent a promising prospect for the future \cite{statista2018}\cite{statista2018a}. 

Wearables are helping end-users to reach a true pervasive and context-aware environment at reduced costs \cite{7854185}. Their effectiveness to sense users' body signals and the environment has been turned these devices into must-have products within certain contexts \cite{juniper2018}. Medical and healthcare wearable devices are frequently used to continuously monitor users' biometric signals, being one of the most common uses of these products \cite{pwc2018}. In addition, wearable solutions are being used in entertainment, sports, by first aid responders and in many other areas. 

Despite the number of solutions and their different applications, wearables on the market sometimes differ from those developed in academic studies. It is common to establish a connection between these two contexts once the market dictates the path in which investigations should go, and vice-versa. The former frequently follows a conservative path, while scientists focus on the freedom to create and try out new concepts and designs. These two approaches allow the exploration of new solutions through multidisciplinary concepts, expanding the horizon of applications and solutions.

However, when considering only academic studies, there is a lack of papers revising and understanding the research’s general trends and impacts on wearable computing. Developing a precise scenario of this area is essential to map the way it evolves, as well as to determine the next steps. Reviews focused on wearables specific to various sub-areas have been presented \cite{7574303}\cite{Swetha:2016:SWC:2980258.2980305}\cite{8263142}, but they do not cover a broad area. There are also papers providing a more general view regarding wearables \cite{Berglund:2016:SHS:2971763.2971796}\cite{7993011}, but do not focus on recent research trends, and thus contribute little to understanding and quantifying current and future efforts to further develop this area.

This article uses a Systematic Review of the Literature (SRL) approach to extract a set of data regarding recent research on wearable computing. Such methodology provides a formal protocol by applying a series of well-defined and reproducible steps, allowing it to be audited by independent peers. This paper aims to provide a general and broad view regarding the trends on wearable computing in academic studies, discussing the results of an SRL, and identifying the current state of the art. The primary intent is to provide a general view of what is being investigated regarding research focus, applicability, functionalities, hardware components, and connectivity. The big picture presented by this work also helps to understand the current scenario on wearables research, expanding the actual investigation focus and possibly indicating the next steps in wearables evolution. In this manner, the authors surveyed a relevant set of papers that describe wearables solutions spanning the past ten years. The following points summarize this paper’s most relevant contributions:
\begin{itemize}
    \item A broad review of wearable computing studies and their applicability; 
    \item A comprehensive and systematic classification of wearables solutions based on a set of constraints;
    \item Identification of main research focus addressed by application area as well as the most relevant technologies used by the devices; 
    \item A general analysis regarding the nature of published work considering wearable devices.
\end{itemize}

Results presented at the end of this paper point to a set of exciting trends. The SRL presents a general view of how the wearables-related studies are scattered around the globe and the impact each country has. In addition, results also provide relevant data of the most common application areas, problems, and technologies used pertaining to wearables research. It is also possible to verify the most trending hardware components and networks used by scientists when proposing new wearable solutions. 

The remainder of this work is divided as follows: Section \ref{sec:wearable_computing} provides a general view of the wearable computing research scenario over the past ten years and provides an original classification of wearables based on a set of specific parameters. Section \ref{sec:related_works} compares the results reached by related studies and how they differ from the research presented by this SRL. Section \ref{sec:work_methodology} presents the methodology used to execute the review, describing the steps as well as the related research questions. Section \ref{sec:results_and_analysis} outlines the main and most important results obtained. Finally, Section \ref{sec:concluding_remark_and_future_works} presents the main conclusions and discusses further work. 
\section{Wearable Computing Research}\label{sec:wearable_computing}
Wearables commonly describe electronic computing devices that can be attached to the human body, aiming to sense specific variables or help the end-user during everyday tasks. They may vary from glasses, watches, strips, hats, exoskeletons, vests, shirts, bags, independent sensors, and any other type of device that a user can wear. 

Despite the increasing number of wearable models and types, this paper does not focus on describing each one, as this task has already been accomplished by other studies \cite{7993011}\cite{Swetha:2016:SWC:2980258.2980305}\cite{7813545}. Instead, this article’s primary intent is to outline the motivations and contexts in which studies on wearables devices have been applied. For this reason, in addition to a historical overview, this section describes a more general classification of wearables, where several parameters are taken into account. 

\subsection{Historical context}
According to the definition of wearables provided in \cite{Dunne:2007:PEW:1240624.1240674}, these devices have been developed since the end of the 1960's, as one of the first HMDs (Head-mounted Displays) \cite{Sutherland:1968:HTD:1476589.1476686}. Since then, there have been three different periods of wearables development \cite{Ariyatum:2005}. During these periods, other proposals and significant results have arisen, encompassing improved design \cite{232473}\cite{Mann:1996:SCS:232014.232021}, development, and evolution of the  wearables field. From HMDs and initial prototypes until today, there has been a ubiquitous set of devices applied to different contexts, e.g., vests, glasses, bands, smartwatches, medical patches, and so on. 

Currently, wearables have reached a level of sophisticated solutions that can embed simple sensors and visually complex components and algorithms. Given this scenario, current and future research may need to explore how the devices can further evolve. For this reason, a more in-depth categorization of novel solutions can help to clarify and identify trends that will guide the next improvements.

\subsection{General classification}
Unlike existing products on the market, research on wearable computing has the freedom to propose and test different concepts and approaches. When analyzing recent investigations, it is possible to verify that a heterogeneous environment introducing various solutions exists. This subsection assigns different classifications to cover all types of wearables presented in the literature. The different classes listed here are further discussed in (Section \ref{sec:results_and_analysis}) and are used to classify and categorize the devices considered by this SRL. 

\subsubsection{Wearable type and body location}\label{subsubsec:wearable_location_and_type}
As presented by \cite{Berglund:2016:SHS:2971763.2971796}, perhaps the most reasonable way to classify wearables is by considering where they are attached to the body. Given the number of existing studies/solutions, a more simplistic way to index them is by considering the human body as three main parts. This classification, in addition to being simple, has no ambiguity. Table \ref{table:wearable_type_body_location} presents a description regarding device types and the location where they are commonly attached to the user’s body. 

\begin{table}[!htbp]
\caption{Wearable Type and Body Location}
\begin{center}
\begin{tabular}{ m{6em} | m{5cm} } 
\hline
\multicolumn{1}{c}{\textbf{Body location}} & \multicolumn{1}{c}{\textbf{Description}} \\ 
\hline
\hline
\textbf{Head} &  Devices that are attached to the user’s head, such as HMDs, glasses, brain-computer interfaces, helmets, hats, and headphones. \\ 
\hline
\textbf{Trunk} & Any wearable covering the trunk part of the body, e.g., shirts, patches, shoulder-worn devices, and belts. \\ 
\hline
\textbf{Limbs} &  Wearable accessories attached to users' arms, hands, legs or feet, such as wristbands, smartwatches, jewels, and finger-worn devices.\\ 
\hline
\end{tabular}
\end{center}
\label{table:wearable_type_body_location}
\end{table}

\begin{figure}[ht]
  \centering
  \includegraphics[scale = 0.45]{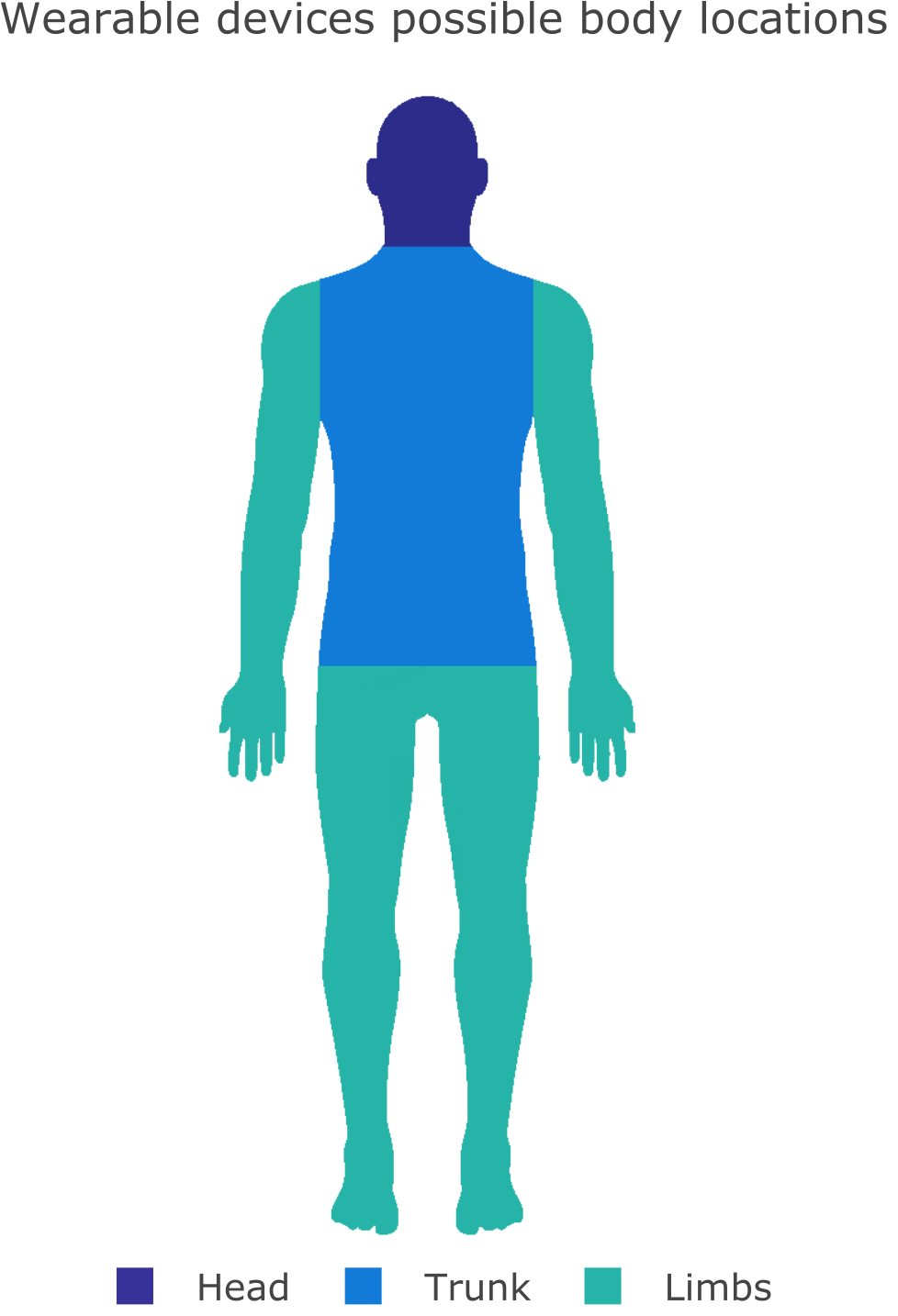}
  \caption{Possible body locations for wearable devices.}
  \label{fig:human_body}
\end{figure}

It is also valid to consider that solutions apply to two or even three body parts. For instance, a whole-body vest characterizes a solution attached to the head, trunk, and limbs at the same time. Figure \ref{fig:human_body} presents a visual representation related to the body locations where wearables can be attached.

\subsubsection{Applicability}\label{subsubsec:applicability}
Another relevant way to classify wearables is by taking their applicability into account. The applicability of a wearable solution refers to the area where it is applied. Although many devices on the market focus on healthcare/medicine and entertainment, studies are now being carried out in several other different contexts. Table \ref{table:wearable_applicability} presents a comprehensive set of areas where fully-functional devices and prototypes are employed. 

\begin{table}[!htbp]
\caption{Wearable devices main applicability areas}
\begin{center}
\begin{tabular}{ m{5em} | m{5cm} } 
\hline
\multicolumn{1}{c}{\textbf{Applicability area}} & \multicolumn{1}{c}{\textbf{Description}} \\ 
\hline
\hline
\textbf{3D modeling} &  Wearables that help to model, test, and interact with 3D structures and representations of the real world \cite{Hoang:2009:AIP:1862703.1862705}. \\ 
\hline
\textbf{Agriculture} & Wearables helping sensing data or physically actuating/supporting some aspect of agriculture tasks \cite{7098714}. \\ 
\hline
\textbf{Artistic} & Devices that can be worn and used as artistic apparels and interventions \cite{Pailes-Friedman:2015:BKA:2800835.2809435}.\\ 
\hline
\textbf{Civil construction} & Solutions that can be used to support heavy or light activities in civil construction \cite{4648016}.\\ 
\hline
\textbf{Communication} & Research proposing the use of wearables to ease or increase communication and information sharing between two or more people \cite{Bace:2017:HWA:3041164.3041203}\cite{Yataka:2011:CAP:1982185.1982274}.\\ 
\hline
\textbf{Cooking} & Wearables that can be used in the context of a kitchen, helping the end-users to perform activities such as cooking or reading a recipe \cite{Matsubara:2015:ERT:2800835.2804333}.\\ 
\hline
\textbf{Driving} & Devices that can help end-users during driving, increasing safety by focusing on the road \cite{Liu:2015:TDU:2753509.2753518}.\\ 
\hline
\textbf{Education} & Solutions aimed to help teachers and general educators to increase the effectiveness of their classes. It can include solutions used by students to receive more contextualized information or boost their attention during classes \cite{Quintana:2016:KWE:2851581.2892493}.\\ 
\hline
\textbf{Entertainment} & Devices inside this applicability area are commonly developed for entertainment purposes (games, movies, music, and so on) \cite{Chan:2015:CWS:2702123.2702464}.\\ 
\hline
\textbf{Environment/ objects and individual sensing} & Wearables used to sense a user’s surrounding context. For instance, devices helping users with visual limitations or other types of context-aware solutions \cite{Ugulino:2015:LIW:2750858.2807541}.\\ 
\hline
\textbf{Fabric} & Devices focused on the development or implementation of new types of wearable fabrics \cite{7525995}.\\ 
\hline
\textbf{Healthcare/ Medicine} & Solutions focusing on healthcare and medicine areas. This encompasses devices aiming to sense general bio-signal data from end-users \cite{Lorincz:2009:MWS:1644038.1644057}.\\ 
\hline
\textbf{Maintenance} & Wearables used to help during maintenance tasks \cite{7028005}.\\ 
\hline
\textbf{Military} & Devices that fall within the military context, helping users directly in the battlefield or in other tasks \cite{5415569}.\\ 
\hline
\textbf{Navigation} & Solutions that can support users during indoor or outdoor movement \cite{Dancu:2015:MNU:2785830.2785876}.\\ 
\hline
\textbf{Security/Rescue} & Devices used to increase the security and safety levels in any environment. It encompasses solutions providing support to security and rescue operations aside from their use by first responders \cite{Chan:2016:EWS:2992154.2996880}.\\ 
\hline
\textbf{Sports} & Devices centered on sports applicability. Solutions to increase athletes’ performance or real-time data monitoring \cite{Schuldhaus:2016:YPM:2971763.2971772}.\\ 
\hline
\textbf{Textile} & Wearables created with focus on textiles. Devices within this category can be used as a basis for other, more complex solutions \cite{Kazemitabaar:2015:MTW:2771839.2771883}.\\ 
\hline
\textbf{Others} & Although the previous list of applicability categories is extensive, there remain other wearables without direct association to the above, e.g., wearables that focus on electronic devices interaction, geology, and mining environments \cite{Delabrida:2016:BWG:2903267.2903275}\cite{Guo:2015:AWC:2702613.2732755}.\\ 
\hline
\end{tabular}
\end{center}
\label{table:wearable_applicability}
\end{table}

In additon to the provided list, the same device may be classified as associated with two or more different areas.

\subsubsection{Research focus}\label{subsubsec:problems}
The applicability areas, as presented in subsection \ref{subsubsec:applicability}, have a close relationship with the challenge addressed by each wearable. Every challenge outlined by this subsection by Tables \ref{table:wearable_research_focus} and \ref{table:wearable_research_focus_II} can be associated to an applicability area. In this manner, the related solutions can also be classified or categorized according to the issue they aim to solve. Below, is a list of the most common challenges tackled by the articles reviewed in this paper.

\begin{table}[!htbp]
\caption{Main topics addressed by wearables research}
\begin{center}
\begin{tabular}{ m{6em} | m{5cm} } 
\hline
\multicolumn{1}{c}{\textbf{Research problem}} & \multicolumn{1}{c}{\textbf{Description}} \\ 
\hline
\hline
\textbf{Autism therapy/support} & Research aiming to increase autism support and provide better therapy alternatives using wearables \cite{Washington:2016:WSI:2851581.2892282}. \\ 
\hline
\textbf{Brain-related diseases} & Devices that address improvement for detection and treatment of these diseases \cite{5090966}. \\ 
\hline
\textbf{Breath analysis} & Through breath pattern analysis it is possible to identify diseases or even infer the type of activity being performed by the end-user \cite{6136953}. \\ 
\hline
\textbf{Cognitive assistance} & Devices that help end-users to complete simple and basic daily tasks \cite{Chen:2015:EIE:2753509.2753517}. \\ 
\hline
\textbf{Context awareness} & Solutions aiming to provide ubiquitous devices to help users by providing context-aware information \cite{Yataka:2011:CAP:1982185.1982274}. \\ 
\hline
\textbf{Daily life monitoring} & Devices that help to identify and monitor the user's daily life and habits \cite{7451069}. \\ 
\hline
\textbf{Dementia support/ monitoring/ therapy} & Wearables that aim to improve the effectiveness of dementia support, monitoring, or therapy \cite{7170674}. \\ 
\hline
\textbf{General diseases detection} & Unlike other more specific problems listed here, this topic involves the detection process of any disease \cite{Mokaya:2015:MVB:2750858.2804258}. \\ 
\hline
\textbf{Energy efficiency} & Devices focusing on energy efficiency improvement in wearables through energy expenditure reduction, energy harvesting, or any other related technique \cite{6151376}. \\ 
\hline
\textbf{Eye tracking} & Wearables addressing how to perform resources-efficient and precise eye identification and tracking \cite{7169792}. \\ 
\hline
\textbf{Fall detection} & Devices focusing on detecting a user’s falling (or pre-falling)  action\cite{Hiyama:2015:BPW:2800835.2800907}. \\ 
\hline
\textbf{Gait assistance/ support/ tracking} & Wearables solutions addressing the problem of gait detection, assistance, support, and tracking \cite{Mazilu:2015:WAG:2744352.2701431}. \\ 
\hline
\textbf{Gaze tracking} & Closely related to eye tracking, this topic involves wearables that focus specifically on gaze direction identification and tracking \cite{Rantala:2015:HFG:2800835.2804334}. \\ 
\hline
\end{tabular}
\end{center}
\label{table:wearable_research_focus}
\end{table}

\begin{table}[!htbp]
\caption{Main topics addressed by wearables research (cont.)}
\begin{center}
\begin{tabular}{ m{6em} | m{5cm} } 
\hline
\multicolumn{1}{c}{\textbf{Research problem}} & \multicolumn{1}{c}{\textbf{Description}} \\ 
\hline
\hline
\textbf{Human activity recognition/ Activities of daily living} &  Wearables that focus on activities recognition problems, mostly using data retrieved from inertial measurement unit (IMU) sensors. The recognition of a set of activities can provide a map of users' behaviors \cite{Younes:2015:IAW:2800835.2801656}. \\ 
\hline
\textbf{Indoor/outdoor localization or navigation} & Devices proposing different solutions for the problems of indoor/outdoor localization or navigation \cite{Garcia:2015:WCI:2811780.2811959}.\\
\hline
\textbf{Personal energy expenditure}: Solutions aimed to retrieve users' daily personal energy expenditure. This problem can be associated with healthcare applications aiming to estimate the number of calories expended during a day \cite{5447796}. \\ 
\hline
\textbf{Physiological parameters analysis} & Any wearable device aiming to retrieve and analyze physiological data parameters to infer end-users' health \cite{Oertel:2016:MBS:2968219.2968574}. \\ 
\hline
\textbf{Posture and gesture recognition} & Wearables used to gather data from sensors to estimate the current posture or even a gesture executed by the user \cite{Wahl:2015:USE:2800835.2800914}. \\ 
\hline
\textbf{Privacy} & Any device which focuses on providing privacy \cite{Preuveneers:2016:PRH:2851613.2851683}. \\ 
\hline
\textbf{Rehabilitation} & Devices focused on the problem of providing any support to a user’s rehabilitation \cite{Song:2016:RWT:2851581.2890229}. \\ 
\hline
\textbf{Remote monitoring} & Wearables that provide any level of remote monitoring during users' activities \cite{6987303}. \\ 
\hline
\textbf{Sleep staging/monitoring} & Devices focusing on monitoring end-users' sleep quality \cite{Nguyen:2016:IBR:2935643.2935649}. \\ 
\hline
\textbf{Stress management/monitoring} & Solutions aimed to provide a set of approaches to monitor or manage user stress \cite{Klamet:2016:WWS:2948963.2948965}. \\ 
\hline
\textbf{User interf. / User exp.} & Wearables that usually have their focus on enhancing the user interface (UI) or user experience (UX) \cite{Yoon:2016:LUI:3016705.3016746}. \\ 
\hline
\textbf{Visually impaired people support/ mobility} & This class includes devices aiming to increase mobility of visually impaired users \cite{Avila:2015:AWT:2786567.2794311}. \\ 
\hline
\textbf{Others} & Despite the relevant applications listed here, there exist other less-popular problems addressed by the researches. Anorexia therapy \cite{6703792}, color vision impairment \cite{7907202}, general data harvesting \cite{Liu:2017:FCW:3024969.3025072}, drug usage detection/monitoring \cite{Natarajan:2013:DCU:2493432.2493496}, and others could be added to this list. \\ 
\hline
\end{tabular}
\end{center}
\label{table:wearable_research_focus_II}
\end{table}


Figure \ref{fig:wearables_mapping} presents a topology proposal that associates the above two subjects providing a better way to understand how the wearables research focus can be related to each area. Its primary motivation is to highlight where the research focus lies. In a more detailed analysis, it is possible to check that some applicability areas fall within different contexts of research. For instance, healthcare/medicine and environment/individual categories are the most addressed contexts among all the considered ones. This fact highlights the relevant and increasing number of wearables solutions covering these two areas. Other challenges, until now, have a reduced scope, being applied only on specific areas, such as ``Breath analysis'' and ``Context awareness''.  

\subsubsection{Technologies, methods, and techniques}
Wearables focused to address the previously listed challenges frequently take into account several different technologies, computing methods, or techniques. Other computing domains so far use these same resources to solve different problems. However, here they have a different focus, being employed inside wearables-related contexts. Technologies, methods, and techniques described in Table \ref{table:technologies_methods_techniques} are the most common ones considered by this SRL.

\begin{table}[!htbp]
\caption{Technologies, methods, and techniques commonly used by wearables research}
\begin{center}
\begin{tabular}{ m{25em} } 
\hline
\multicolumn{1}{c}{\textbf{Tech./ method or technique}} \\ 
\hline
\hline
General Artificial Intelligence techniques, methods, and algorithms. \\ 
\hline
Augmented Reality and Virtual Reality (AR/VR). \\ 
\hline
Data from Ballistocardiogram (BCG). \\ 
\hline
Brain-computer Interfaces (BCI). \\ 
\hline
Beacons. \\ 
\hline
Bio-impedance. \\ 
\hline
Data from Channel State Information (CSI). \\ 
\hline
Crowdsourcing. \\ 
\hline
Deep Brain Stimulation (DBS) methods. \\ 
\hline
Digital Image Processing techniques. \\ 
\hline
Discriminant Analysis methods. \\ 
\hline
Dynamic Time Warping (DTW). \\ 
\hline
Electrical Impedance Tomography (EIT). \\ 
\hline
Electrodermal Activity (EDA) classification. \\ 
\hline
Electrophoretic Displays. \\ 
\hline
Face Recognition methods. \\ 
\hline
Fast Fourier Transforms (FFT). \\ 
\hline
Graphene material. \\ 
\hline
Haptic technology and applications. \\ 
\hline
Heart Rate Variability (HRV). \\ 
\hline
Histogram of Oriented Gradients (HOG). \\ 
\hline
Human Hybrid Robots (HHR). \\ 
\hline
Internet of Things (IoT). \\ 
\hline
Least Mean Square (LMS) algorithm. \\ 
\hline
Markov Chains. \\ 
\hline
Motion History Images (MHI). \\ 
\hline
Object Recognition (using SIFT/MOPED algorithms). \\ 
\hline
Optical Character Recognition (OCR). \\ 
\hline
Data from Photoplethysmography (PPG). \\ 
\hline
Data from Root Mean Square (RMS). \\ 
\hline
Seismocardiography (SCG). \\ 
\hline
Solar Energy to power devices. \\ 
\hline
Wearable Power Assist Leg (WPAL). \\ 
\hline
Others. \\
\hline
\end{tabular}
\end{center}
\label{table:technologies_methods_techniques}
\end{table}

Items present above are not an exhaustive list of ``all'' technologies, methods, and techniques used by wearable devices. Instead, they represent the most common items used by research to propose wearable solutions. Because of their wide scope, some may contain other relevant subsets. For instance, ``Artificial Intelligence'' may embrace a significant set of algorithms, models, and methods used by related studies (Table \ref{table:artif_intelligence_items}): 

\begin{table}[!htbp]
\caption{Main artificial intelligence topics in wearable research}
\begin{center}
\begin{tabular}{ m{19em} } 
\hline
\multicolumn{1}{c}{\textbf{Algorithm, method, or model}} \\ 
\hline
\hline
Bayes nets and Dynamic Bayesian Networks (DBN). \\ 
\hline
Convolutional Neural Networks (CNN). \\ 
\hline
Decision Trees. \\ 
\hline
Haar Cascade Classifiers. \\ 
\hline
Hidden Markov Models (HMM). \\ 
\hline
K-Means algorithms. \\ 
\hline
Logistic Regression algorithms. \\ 
\hline
Logit Boost algorithm. \\ 
\hline
General Machine Learning methods/algorithms. \\ 
\hline
Nearest-neighbor approaches (including k-NN). \\ 
\hline
Random Forest algorithm. \\ 
\hline
Supporting Vector Machine (SVM). \\ 
\hline
Others. \\ 
\hline
\end{tabular}
\end{center}
\label{table:artif_intelligence_items}
\end{table}

\afterpage{
\begin{figure*}[tp]
  \centering
    \begin{turn}{90}
    \begin{minipage}[!t]{9.7in}
  \centering
  \includegraphics[scale = 0.1]{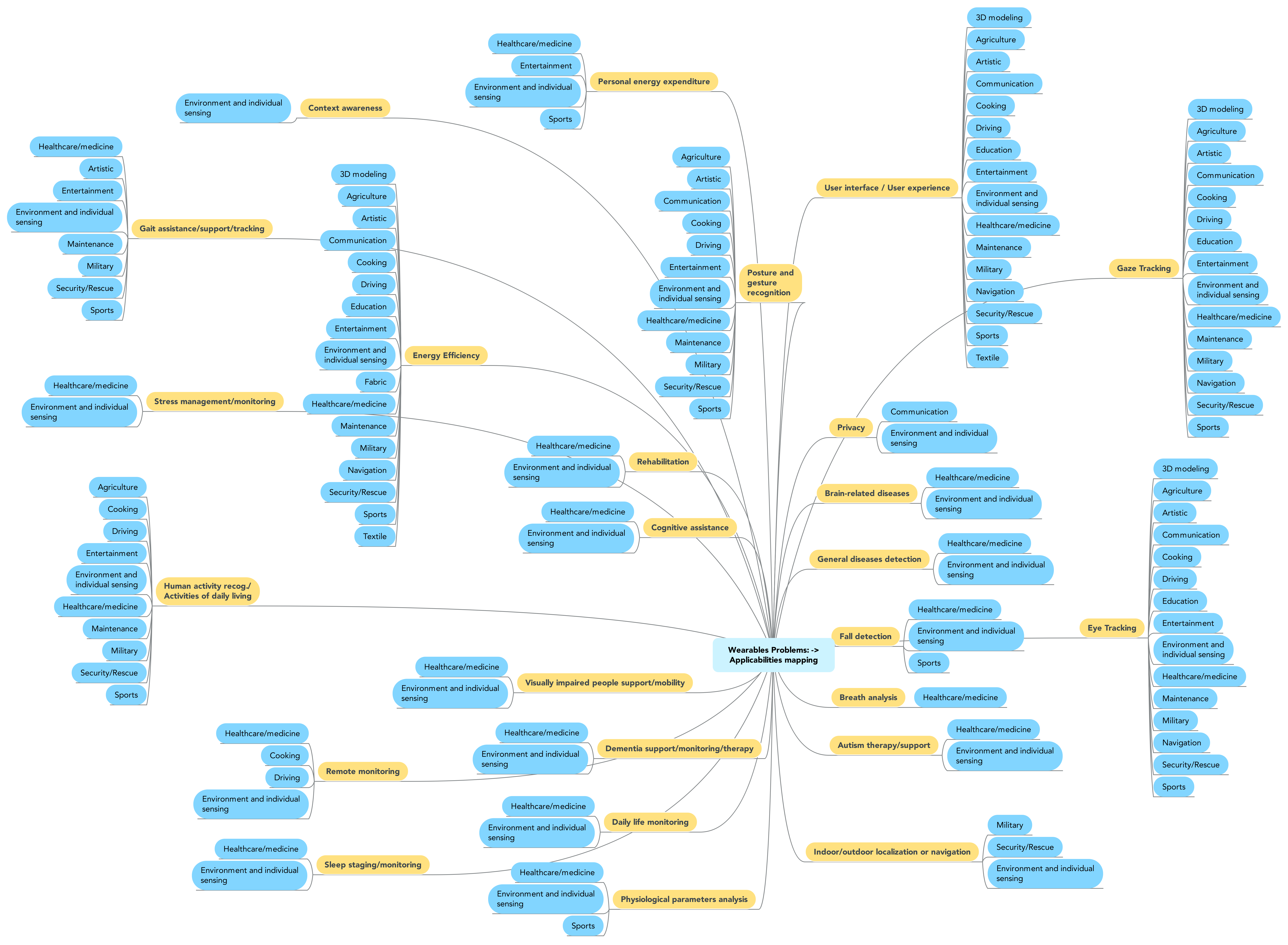}
  \caption{Wearables problems contextualized into each applicability area.}
  \label{fig:wearables_mapping}
  \end{minipage}
  \end{turn}
\end{figure*}
}

\subsubsection{Connectivity \& networking}\label{subsubsec:connectivity}
From the simplest wearables to the most complex ones, numerous solutions are using some network connection type to send or receive information. Indeed, the connectivity degree of these devices dictates the data range. Briefly, they can form a wearable body area network (WBAN) or even be integrated to outside networks sending information to a server in the cloud.  Table \ref{table:network_connectivity_protocols} lists the most common network types and protocols used by wearables included in this SRL. 

\begin{table}[!htbp]
\caption{Network types and protocols used by wearables}
\begin{center}
\begin{tabular}{ m{19em} } 
\hline
\multicolumn{1}{c}{\textbf{Network type / protocol}} \\ 
\hline
\hline
Ant+. \\ 
\hline
Bluetooth. \\ 
\hline
Cellular networks (3G, 4G, 5G, GPRS, GSM, CDMA, TDMA). \\ 
\hline
Near Field Communication (NFC). \\ 
\hline
RFID. \\ 
\hline
Wi-Fi. \\ 
\hline
XBee. \\ 
\hline
Zigbee. \\ 
\hline
\end{tabular}
\end{center}
\label{table:network_connectivity_protocols}
\end{table}


\section{Related Work}\label{sec:related_works}
Research studies in wearable computing have a close relationship with the wearables offered on the market. As soon as products’ numbers increase to indicate popularity, or if complexity is added or identified as necessary, more interest is generated within academic circles. On the other hand, the more research performed, the greater are the number of products arriving on market shelves. This feedback loop increases the solutions on both sides, requiring companies and scientists to have available a comprehensive view of the area.  

Despite the relevant number of surveys on wearable computing, few have focused on providing an in-depth view of the global research being carried out. This section focuses on recent work related to the research areas presented by this review, separating them according to comprehensive scope.

\subsection{General research}
Inside this class fall the general focus related studies covering wearable computing/technology. 

In the paper presented by \cite{7993011}, a classification and survey are presented, with the authors describing commercially available products and research prototype solutions. An excellent approach is taken, focusing on several aspects that influence wearables contexts, such as the current state of the art and the solutions classification of ``Accessories'', ``E-Textiles,'' and ``E-Patches'' within the groups of ``Existing Products'' and ``Research prototypes''. The work also focuses on providing an SRL regarding communication security issues and challenges in energy efficiency, while always considering the wearables environment. Thus, the work is closely related to the one presented by this paper, and they may be considered as complementary, showing initial concepts, general trends, and the current status regarding the wearables area, but with different focus.

In \cite{Shrestha:2017:ODE:3161158.3133837} a general review on wearables is provided, contextualizing them within applicability domains. Emphasis is placed on the security, privacy, and usability subparts, describing possible threats, risks, and defenses of this type of device. The work exhibits good coverage and organization. However, the paper is centered only on security and privacy questions.

The survey presented in \cite{Berglund:2016:SHS:2971763.2971796} focuses on research prototypes and industrial solutions on the market. A historical survey is proposed as a way to understand how the area is evolving and what the current focus is on wearable devices. The review presents a solutions categorization according to the place where the device is attached to the users' body. Furthermore, the work compares historical trending data with the current tendencies. In comparison with our work, it lacks several types of information. As can be verified from the results section (Section \ref{sec:results_and_analysis}), it is possible to contextualize the wearables trending data not only according to the applicability area and body location but also by considering a set of other parameters that will help to better clarify the research directions.

In \cite{7813545} the authors present another survey, classifying wearable devices according to the scope where they are applied: ``Assistive'', ``Workplace'', ``Healthcare'', and ``Consumer Products''. Despite its vast scope, the work mainly focuses on devices already on the market, with a limited number of products being analyzed. When compared to the work described here, it is possible to verify that a different approach was given when analyzing wearable devices by prioritizing the research area. In addition, this research has used an increased amount of research studies as a means to increase the results’ confidence level.  

\subsection{Specific research}
Unlike the previous subsection, section centers on reviews with specific focus within the wearables field. Instead of providing a more general classification, related work described below focuses on reviewing distinct subparts or specific components that integrate with the wearable computing area.

Within wearable computing, there are several different approaches. One is the study of how to sense human body characteristics aiming to discover user behavior and activities. In \cite{Lioulemes:2016:SSM:2910674.2910711} the authors reviewed emerging technologies used to sense users' activities and physiological variables. It presented a set of devices dedicated to retrieve information from different areas of the human body, such as glasses, heart rate sensors, blood pressure sensors, and so on. In general, the work focuses on healthcare. Conversely, in \cite{7742959} the authors also present a review focused on wearable sensing, concentrating on activity classification. Different sensors are discussed as well as research covering single and hybrid sensing modalities. The work described here has a broader scope, encompassing both wearable sensing and activity classification while also including other areas.

A survey on wearable medical sensor systems is presented by \cite{7870697}, describing the application, typical architecture, and components that are used to build these systems. A systematic classification is described, dividing the devices into ``Healthcare'', ``Human-computer interaction (HCI)'', ``Security and forensics'', and ``Education'' areas. Notwithstanding, as the scope of the work is limited, it lacks more extensive coverage that may include additional wearables applications and problems, as done by the research presented in this SRL.

In \cite{Blasco:2016:SWB:2988524.2968215} a survey regarding wearable biometric recognition systems is presented. Unlike the previous studies, this paper touches upon an even more specific subject inside wearable computing: biometrics. Information presented by the article helps to define a biometrics taxonomy using wearable sensors. The authors also present concepts of signal processing techniques, data classification, and machine learning algorithms. Although very well executed, the survey presents a more specific range than the results described by this SRL, which aims to understand how the entire wearable devices area has evolved over the past years.

\section{Work Methodology}\label{sec:work_methodology}
\begin{figure}[ht]
  \centering
  \includegraphics[scale = 0.36]{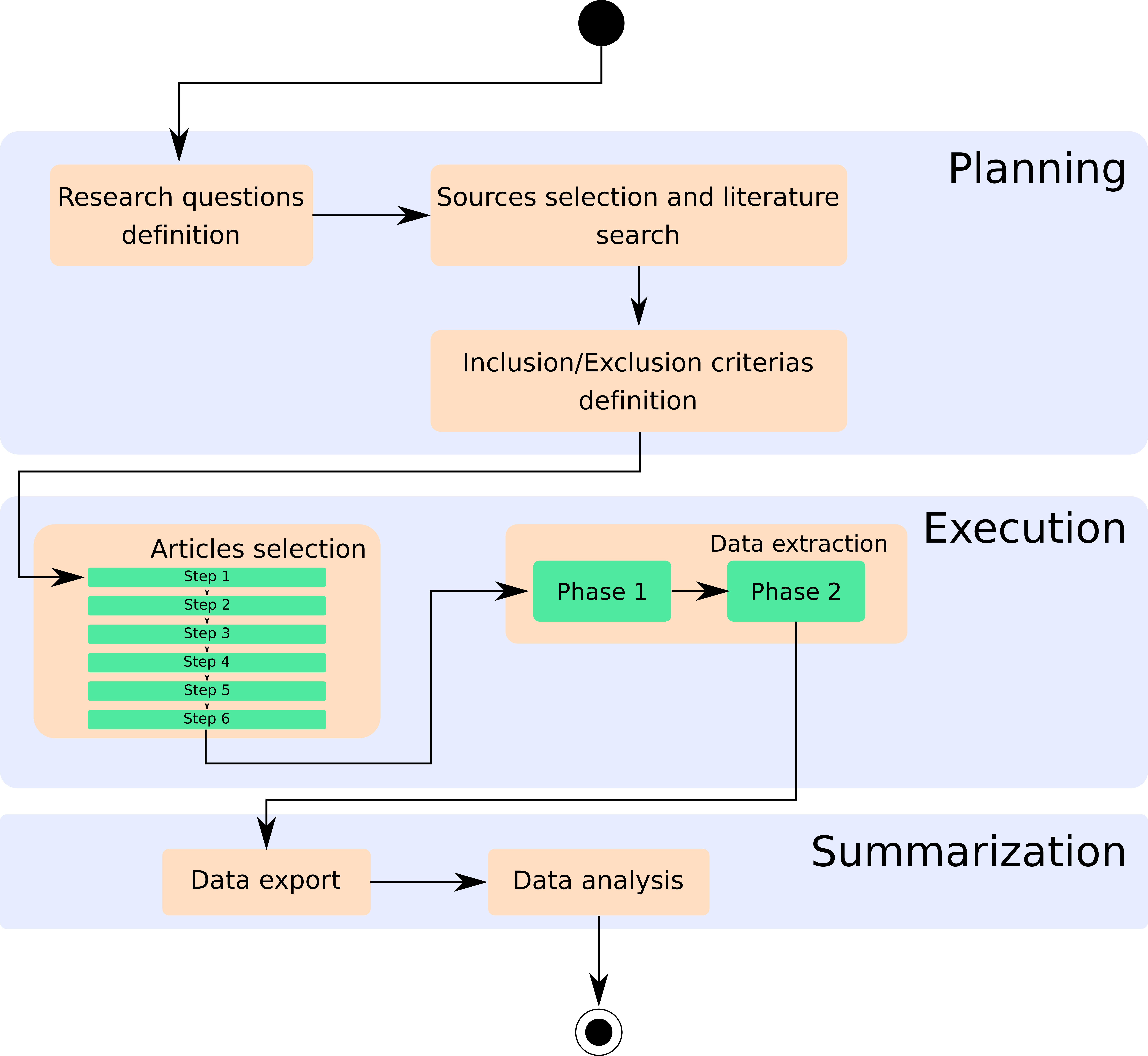}
  \caption{Flowchart showing the inner tasks executed by the three main modules.}
  \label{fig:review_steps}
\end{figure}

An SRL is a way to recognize, compare, and evaluate relevant research in a particular area or context. Data raised by an SRL allows inferring trends and patterns using an auditable and trustable methodology \cite{kit_cha_2007}. The SRL presented in this paper follows the guidelines proposed by \cite{kit_cha_2007}, dividing the work into three main modules: ``Planning'', ``Execution'', and ``Summarization''. During the development, an auxiliary tool called ``State of the Art through Systematic Review'' (StArt) \cite{start2018} was used. Figure \ref{fig:review_steps} details the initial actions and inner tasks taken for each module.

\subsection{Planning}
During this phase, the SRL protocol was designed focusing on its main objective: ``identify and categorize recent proposals of wearable computing devices with their applications, applicability areas, and technologies available in the literature''. To fulfill this objective, two main research questions were proposed:

\begin{enumerate}
    \item \textbf{Main Research Question (RQ1)}: What are the main areas in which recent wearable computing device studies have been applied?
        \begin{itemize}
            \item \textit{Motivation}: Understand the context and applicability areas where wearable research studies have been carried out. This information will allow scientists to target their current and future works better;
            \item \textit{Expected results}: Produce a measurable conclusion regarding the top trending application areas for wearable computing devices.
        \end{itemize}
    \item \textbf{Secondary Research Question (RQ2)}: What types of features are mostly used by wearable computing device studies?
        \begin{itemize}
            \item \textit{Motivation}: Discover the trending topic technologies currently being considered to craft wearable devices investigations.
            \item \textit{Expected results}: Provide a clear separation of wearable computing solutions so that it will be possible to categorize them according to the considered features, technologies, and methods being used in the academic environment when researching wearables subjects.
        \end{itemize}
\end{enumerate}

\subsubsection{Source selection and literature search}\label{subsubsec:literature_search}
Through an initial search of potentially relevant sources, the authors have identified a set of repositories indexing articles related to the wearable environment. The following electronic databases were taken into account: ACM Digital Library\footnote[1]{https://dl.acm.org/}, IEEE Xplore Digital Library\footnote[2]{http://ieeexplore.ieee.org/Xplore/home.jsp}, Springer Link\footnote[3]{https://link.springer.com/}, Science Direct\footnote[4]{https://www.sciencedirect.com/}, Scopus\footnote[5]{https://www.scopus.com/home.uri}, and Engineering Village\footnote[6]{https://www.engineeringvillage.com/home.url}. Articles written in the English language were identified within these repositories using the arrangement of wearables-related keywords presented in Table \ref{table:search_keywords}.

\begin{table}[tb]
\caption{Wearables-related keywords used to search article repositories}
\begin{center}
\begin{tabular}{ m{20em} } 
\hline
\multicolumn{1}{c}{\textbf{Search term}} \\ 
\hline
\hline
\emph{ARTICLE\_TITLE(wearable*) AND ARTICLE\_ABSTRACT(activity OR augment* OR eye OR tracking OR gesture OR haptic OR head-mounted OR smartwatch OR virtual OR wrist-worn OR intelligent OR finger OR pervasive OR embedded OR hardware OR software OR medical OR camera OR recognition OR application OR applicability OR context OR helmet OR glasses OR display OR electronic OR sensor OR motion OR mobile OR hand OR cloth* OR smart OR personal OR screen)} \\
\hline
\end{tabular}
\end{center}
\label{table:search_keywords}
\end{table}

This keywords-based search was applied to each electronic database, retrieving articles published between 2008/Jan/01 and 2017/April/19. This time interval was considered by the authors as being sufficient to provide a high-quality view of the recent research being carried out in the wearable computing devices field.

\subsubsection{Inclusion and exclusion criteria}
As soon as the studies were gathered from the literature search, a set of inclusion/exclusion rules were applied to them as a means to refine the number and increase the quality of obtained articles. Initially, a study was considered suitable for this SRL if it met the following inclusion criteria:

\begin{itemize}
    \item \textbf{Inclusion criteria:}
    \begin{itemize}
        \item Full articles published between 2008/Jan and 2017/April/19 in English. The language restriction was applied because the authors understand that the research will have a more significant impact when written in this language;
        \item Studies presenting any novel wearable device describing its real-world applicability and motivation problem; and
        \item Articles describing any novel wearable device and its internal components, technologies, methods, or hardware components.
    \end{itemize}
\end{itemize}

Given the number of retrieved articles from the ``literature search'' stage, a set of exclusion rules were also crafted. Then, any article that met at least one of these rules was removed from the SRL.

\begin{itemize}
    \item \textbf{Exclusion criteria:}
    \begin{itemize}
        \item Articles without an explicit title or author were removed;
        \item Papers without the word ``wearable*'' in the title were not considered. Although this restriction was already considered by the keyword search, some digital repositories were apparently unable to comply with this constraint; 
        \item Studies in any language other than English were eliminated;
        \item Opinions, reports, talks, and other non-scientific studies were also not considered;
        \item Short-papers, demos, abstract-only articles, comments, and other works with less than three pages were removed from the SRL. Such decision was taken because of the reduced amount of information available from these papers;
        \item Proceedings documents were removed as they index articles already listed by the digital repositories independently;
        \item Systematic reviews, reviews, and surveys were excluded. As in the proceedings, this type of study commonly indexes other research. Thus, removing then avoids analyzing duplicated work;
        \item Retracted articles;
        \item Studies not available in electronic format were not analyzed. Although being returned by the digital repositories searches, some articles are no longer available in an electronic format;
        \item Articles with some digital access restrictions (i.e., that could not be digitally retrieved) were not considered;
        \item Papers whose titles did not describe a wearable device/solution, such as antennas, data analysis, standard specifications, and so on;
        \item Articles in which the Abstract did not mention a wearable device/solution being proposed or described;
        \item All studies with two or fewer pages. As is common with most demonstrations and posters, papers with two or fewer pages commonly do not have enough room to properly describe the work methodology or provide a full results section covering a wearable device, its internal components, and applicability;
        \item Works that did not describe their methodology or results were not provided; and
        \item Out of context studies. Papers focusing on aspects other than those considered by this SRL were discarded.
\end{itemize}
\end{itemize}

\subsection{Execution}\label{subsec:execution}
In this step, data raised by the two previous steps were used to search for specific and high-quality studies that precisely fit the SRL parameters. Once the sources were selected, the queries performed, and the inclusion/exclusion criteria defined, it was necessary to select which works were better qualified to be considered for this review.

\subsubsection{Articles selection}
\begin{figure}[ht]
  \centering
  \includegraphics[scale = 0.63]{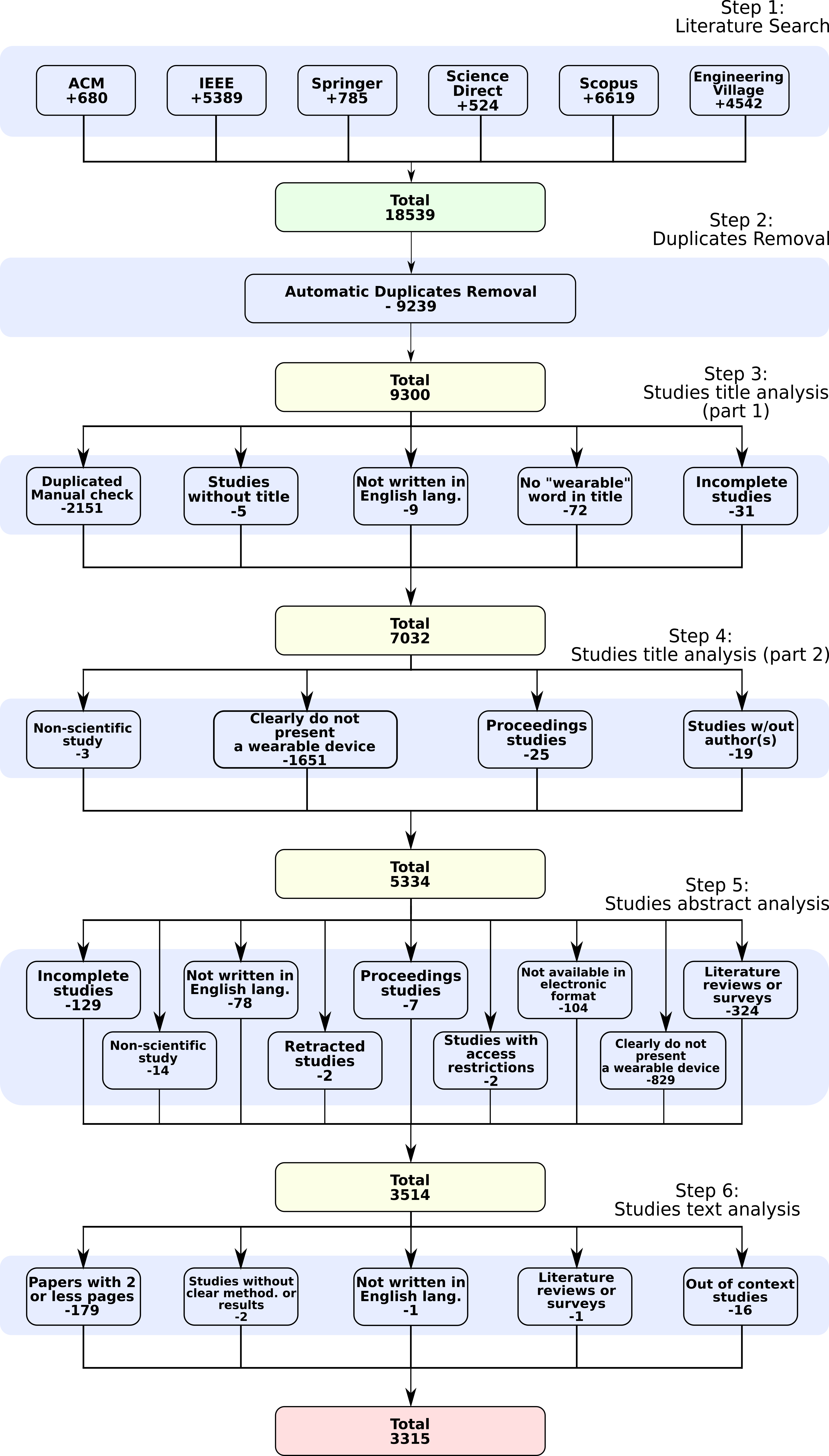}
  \caption{Flowchart describing the steps taken during article selection.}
  \label{fig:review_flow}
\end{figure}

After the literature search presented in subsection \ref{subsubsec:literature_search}, all the retrieved papers were subjected to the previously described inclusion/exclusion criteria. Owing to the number of results, the procedure was divided into five steps, as shown in Table \ref{table:selection_phase_steps}. Figure \ref{fig:review_flow} presents a detailed flowchart of what actions were taken in each step. From an initial collection of 18,539 articles, 3,315 articles were ultimately selected.

\begin{table}[tb]
\caption{Steps taken during article selection}
\begin{center}
\begin{tabular}{ m{11em} | m{5cm} } 
\hline
\multicolumn{1}{c}{\textbf{Step Title}} & \multicolumn{1}{c}{\textbf{Description}} \\ 
\hline
\hline
Step 1: Literature search &  Search in the digital repositories looking for articles that meet the constraints defined in subsection \ref{subsubsec:literature_search}.  \\ 
\hline
Step 2: Duplicates removal & Automatic verification and removal of duplicated papers using StArt tool. \\ 
\hline
Step 3: Studies title analysis (part 1) &  Manual verification and removal of duplicated articles. Additionally, papers titles were verified looking for: studies without a title, works not written in English language, papers whose title does not contain the ``wearable*'' word, and incomplete studies.\\ 
\hline
Step 4: Studies title analysis (part 2) &  Second round of paper title analysis. This time studies fitting at least one of the following exclusion criteria were removed from the SRL: Non-scientific studies, works that do not present a wearable device/solution, proceedings, and articles without author(s).\\ 
\hline
Step 5: Studies Abstract analysis & Verification of studies Abstract information. Works whose Abstract fit at least one of the following exclusion criteria were removed: Incomplete studies, works not written in English, proceedings studies, articles not available in electronic format, literature reviews or surveys, non-scientific studies, retracted papers, works with some digital access restrictions, and articles whose Abstracts clearly do not mention a wearable device/solution.\\ 
\hline
Step 6: Papers full-text verification & This step searched for specific information throughout the paper, mainly checking the Introduction section (or its equivalent), Methodology, and Results, removing articles that lacked some of these parts. Moreover, for works addressing literature reviews and surveys, papers with two or fewer pages, works not written in English, and out of context studies were removed.\\ 
\hline

\end{tabular}
\end{center}
\label{table:selection_phase_steps}
\end{table}

\subsubsection{Data extraction}
After the literature search and articles selection (using defined inclusion/exclusion criteria), data were extracted from the papers. The extraction form used by this review looks for the information presented in Table \ref{table:extracted_data}, where the first column outlines the data type and the second column the possible values. A short description of what it means is also provided. Owing to the large number of studies, the data extraction step was divided into two phases:

\begin{itemize}
    \item \textbf{Phase 1:} Approximately 10\% (338) of the studies were manually analyzed. Introduction, Conclusion and Methodology sections were read, while the remaining text was screened to extract the data presented in Table \ref{table:extracted_data}. For each type of data being extracted, a set of keywords was also collected. These keywords were further used in the next step to automatically retrieve the remaining papers’ (2,977) contents using a specially-developed mining script.
    
    \item \textbf{Phase 2:} Using a specific script and a set of keywords collected during the previous phase, the data listed in Table \ref{table:extracted_data} (except those in the highlighted rows) were automatically mined and extracted from each of the remaining 2,977 articles. This phase was necessary owing to the large number of articles to be analyzed after the filter applied in Step 6. 
\end{itemize}

\begin{table*}[tb]
\caption{Summary of data extracted from selected articles}
\begin{center}
\begin{tabular}{ m{11em} | m{14cm} } 
\hline
\multicolumn{1}{c}{\textbf{Data title}} & \multicolumn{1}{c}{\textbf{Description / Possible values}} \\ 
\hline
\hline
\textbf{Basic article metadata} & Title, author(s), affiliation(s), affiliation country(ies), publication year, and publication vehicle (journal, conference, book, chapter, ...) \\ 
\hline
\rowcolor{Gray}
 \textbf{Study type} & Studies were classified according to their type among the following options: Validity study, case study, work in progress, interview, initial study, finished work.   \\ 
\hline
\rowcolor{Gray}
 \textbf{Does the work focus on a wearable device?} & Verify if (yes/no) the wearable device is the main research object, i.e., the focal point for each paper. \\ 
\hline
\rowcolor{Gray}
 \textbf{Does the work include a prototype description?} & Verify if (yes/no) the article describes a wearable device prototype as a means to validate the proposition. \\ 
\hline
\rowcolor{Gray}
 \textbf{Wearable device description level} & Measure how detailed the wearable device description is. Possible values are: (0) Wearable solution is not described; (1) Solution device is barely cited through the text; (2) Device is mentioned in the text and presented through images; (3) Wearable prototype and its implementation are described by the text and figures; and (4) Wearable prototype is deeply described, presenting the hardware used to design and develop the solution. \\ 
\hline
\rowcolor{Gray}
\textbf{Is the wearable prototype a complete solution?} & Check if (yes/no) the proposed solution is a fully-functional wearable device or just a part/module of a more complex solution, e.g., a sensor, an improvement, and so on.\\
\hline
\textbf{Wearable location} & Extract the human body part where the wearable device is attached: Head, trunk, or limbs. A consistent description for the scope of each part can be found in subsection \ref{subsubsec:wearable_location_and_type}.\\ 
\hline
\textbf{Wearable type} & Classify the proposed solution into one of the following types: wristband, smartwatch, exoskeleton, helmet, necklace, sub-vocal sensor, belt, glass, chest device, vest, glove, in-ear, finger-worn, shoe sensor, goggles, strip, halo, headphone, or independent sensor(s) attached to different parts of the users' body, bag, jewel, hat, knee, and others.\\
\hline
\textbf{Applicability area} & Contextualize the wearable device into one of the following applicability areas: 3D modeling, agriculture, artistic, civil construction, cooking; education, entertainment, environment/objects and individual sensing, fabric, healthcare/medicine, maintenance, military, navigation, security/rescue, sports, textiles, and others. A full description of what each applicability area covers can be found in subsection \ref{subsubsec:applicability} of this paper. \\
\hline
\textbf{Research Focus} & Wearables solutions are categorized according to their research focus: autism therapy/support, brain-related diseases, breath analysis, cognitive assistance, context awareness, daily life monitoring, dementia support/monitoring/therapy, general diseases detection; energy efficiency; eye tracking; fall detection; gait assistance/support/tracking, gaze tracking, human activity recognition/activities of daily living, indoor/outdoor localization or navigation, personal energy expenditure, physiological parameters analysis, posture and gesture recognition, privacy, rehabilitation, remote monitoring, sleep staging/monitoring, stress management/monitoring; user interface / user experience, visually impaired people support/mobility, and others. A more detailed description regarding each problem can be found in subsection \ref{subsubsec:problems} of this article. \\
\hline
\textbf{Technologies, methods and techniques} & Devices were associated and classified according to the technology, method, or technique using them. The following possible values were considered according to the item relevance: General Artificial Intelligence techniques, methods and algorithms, Augmented Reality and Virtual Reality (AR/VR), data from ballistocardiogram (BCG), brain-computer interfaces (BCI), Beacons, bio-impedance, data from channel state information (CSI), crowdsourcing, deep brain stimulation (DBS) methods, digital image processing techniques, discriminant analysis methods, dynamic time warping (DTW), electrical impedance tomography (EIT), electrodermal activity (EDA) classification, electrophoretic displays, face recognition methods, fast Fourier transform (FFT), graphene material, haptic technology, heart rate variability (HRV), histogram of oriented gradients (HOG), human hybrid robots (HHR), internet of things (IoT), least mean square (LMS) algorithm, Markov chains, motion history images (MHI), objects recognition (using SIFT/MOPED algorithms), optical character recognition (OCR), data from photoplethysmography (PPG), data from root mean square (RMS), seismocardiography (SCG), solar energy, wearable power assist leg (WPAL), and others.  \\
\hline
\textbf{Connectivity \& networking} & Extract connectivity and network-related information from each reviewed work. As presented in subsection \ref{subsubsec:connectivity}, the possible values are: ANT+, Bluetooth, cellular networks (3G, 4G, 5G, GPRS, GSM, CDMA, TDMA), NFC, RFID, Wi-Fi, XBee, and ZigBee.\\
\hline
\textbf{Associated layer} & As presented by \cite{8319168}, wearables can be categorized into five different layers according to their complexity level (Layer 0 being the simplest and Layer 4 considering the most complex). \\ 
\hline
\textbf{Hardware components} & Most common hardware components used to craft wearables prototypes and solutions were retrieved from the selected studies. Some examples are: accelerometer sensors, analog to digital converters, antennas, arduinos, cameras, ECG sensors, EEG sensors, electrodes, Google Glass, global positioning system (GPS), gyroscope sensors, heart rate sensors, impedance analyzers, magnetic sensors, microcontrollers, microphones, printed circuit boards (PCB), PPG sensors, pressure sensors, and others. \\
\hline
\end{tabular}
\end{center}
\label{table:extracted_data}
\end{table*}

\subsection{Summarization}  
After the literature search and articles selection, data was extracted from each of the 3,315 papers using the predefined form presented in Table \ref{table:extracted_data}. The resulting information was automatically exported to a spreadsheet using the support provided by the StArt tool. Data inside the exported spreadsheet was then separately analyzed using specially built scripts. These scripts isolate each information type and correlate them to the work publication year, providing significant results to understand the trending topics of each part. Section \ref{sec:results_and_analysis} presents a deep analysis of each resulting data and presents them with appropriate contextualization. 

\section{Results and Analysis}\label{sec:results_and_analysis}
After the previously described procedures performed in the ``Execution'' and ``Summarization'' parts, the resulting data consists of information from 3,315 valid articles. Results presented in this section analyze the summarized information, separating it into two main groups (Group 1 and Group 2) according to the nature of the information. While Group 1 focuses on presenting information related to the nature of the applications and their characteristics, Group 2 is a compilation of results considering data extracted concerning the proposed wearables.

\subsection{Group 1}
Results presented here cover only manually extracted data, highlighted in Table \ref{table:extracted_data}, from 338 randomly chosen articles. Initially, data compiled here is only concerned about the study’s nature describe the general characteristics of research papers published in the wearables area. The authors took this decision as a way to ease the reader's understanding of this SRL, as due to their nature, this type of data is particularly difficult to be automatically extracted from articles. Furthermore, this initial data also provided a significant set of samples regarding the analyzed articles and the wearable prototypes they propose.

Table \ref{table:group1_results} presents all the 338 papers’ manually extracted data. Resulting information was then grouped into five different sets following the organization outlined by the highlighted rows of Table \ref{table:extracted_data} and is further explained below.

\subsubsection{Study type}
Articles were initially classified according to the study type, considering the following possible types:

\begin{itemize}
    \item \textbf{Finished Works (FW)}: Works whose consistent results are presented following a methodology and address reach objectives inside the listed scope;
    \item \textbf{Work in Progress (WP)}: Articles that describe a work in progress, i.e., part of, or a module that will compose a complete solution;
    \item \textbf{Case Study (CS)}: Papers that focus on a case study;
    \item \textbf{Validity (VL)}: Works that aim to demonstrate the validity of a hypothesis, theory, or physical component/prototype through a valid experiment;
    \item \textbf{Interview (IV)}: Articles focusing on data obtained from interviews with possible users or specific interest groups;
    \item \textbf{Initial Study (IS)}: Papers describing an initial study, which is finished but does not provide a complete solution. 
\end{itemize}

\begin{figure}[ht]
  \centering
  \includegraphics[scale = 0.39]{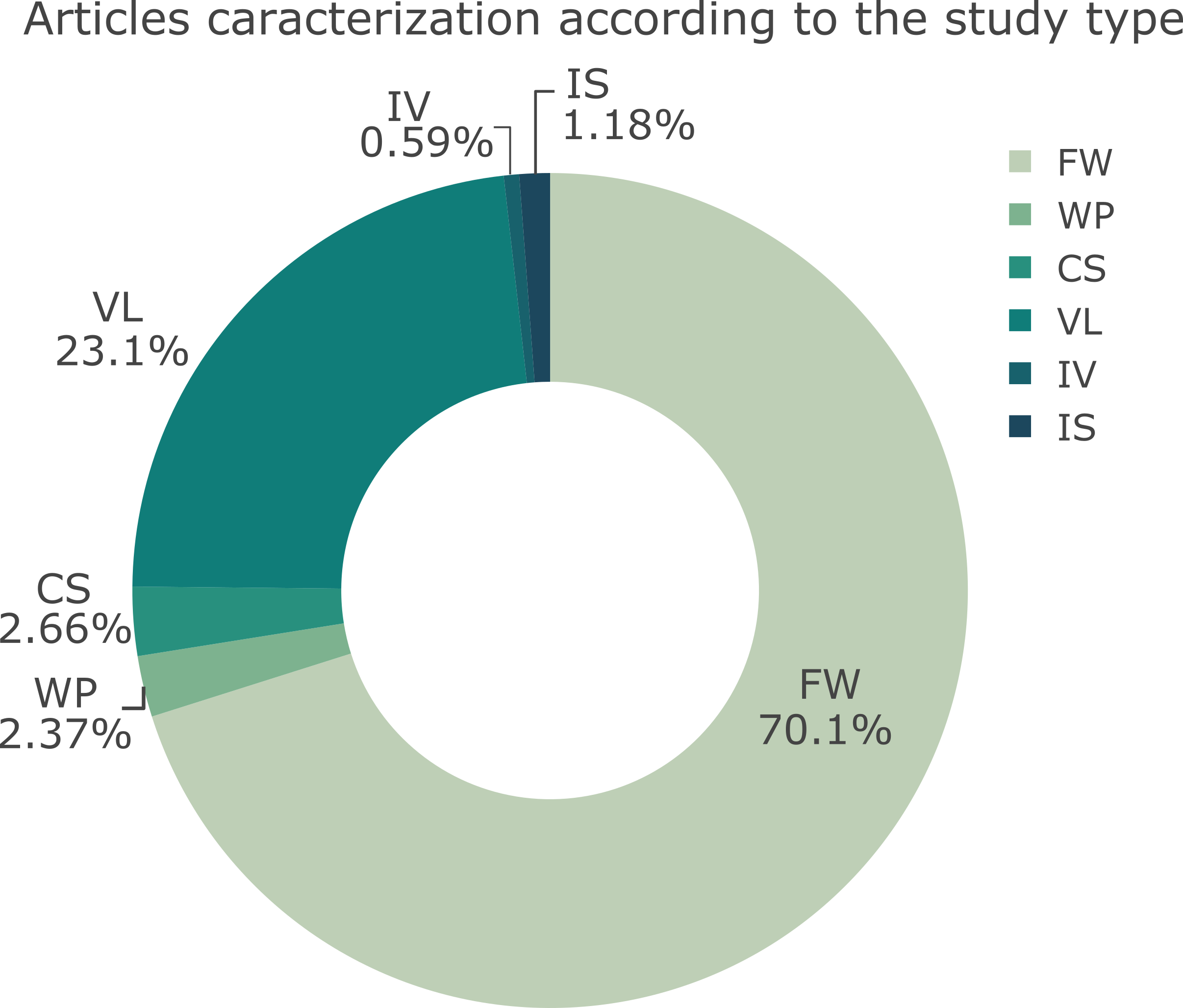}
  \caption{Manually reviewed works characterized according to the study type.}
  \label{fig:study_type}
\end{figure}

Figure \ref{fig:study_type} shows a visual representation of how the articles can be categorized. Most wearables solutions presented in the papers can be classified as ``Finished Works (FW)'', as they have a closed scope and present a functional prototype. Other interesting data relate to the ``Validity (VL)'' studies: A reasonable amount of studies use this approach to evaluate users' perceptions or to validate an idea. ``Case Study (CS)'', ``Work in Progress (WP)'', and ``Initial Study (IS)'' papers represent 6.21\% of the studies, indicating that they have low relevance overall. This result may suggest that scientists focus on publishing finished works rather than disclosing tentative conclusions. Another hypothesis is that few journals/conferences in the area accept ``under development'' articles. Notwithstanding, this theory must be confirmed with a deeper analysis. Finally, only 0.59\% of articles were classified as ``Interview (IV)'', demonstrating a low scientific motivation for publishing this type of study.

\subsubsection{Does the work focus on a wearable device} 
This question is asked to determine whether the wearable device described in the article is the research’s main objective/result. This question was raised as a way to describe the real objectives when assessing wearable computing studies: Are they focused on building wearable devices and/or are they used as a technical solution (or part of one) in other problems/areas?

\begin{figure}[ht]
  \centering
  \includegraphics[scale = 0.27]{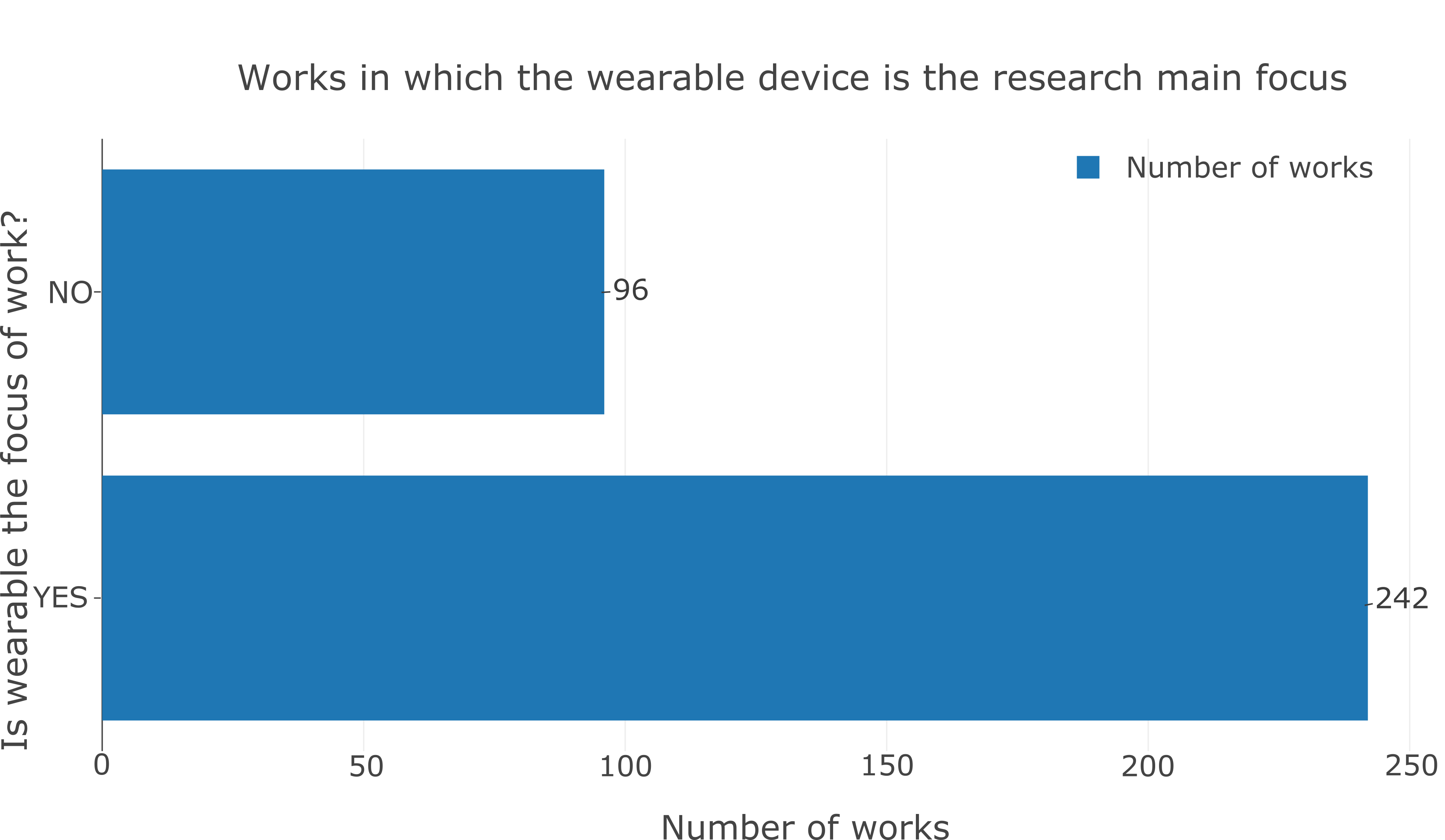}
  \caption{Number of studies in which the wearable device is the main research output.}
  \label{fig:wearable_is_focs}
\end{figure}

The Figure \ref{fig:wearable_is_focs} results indicate that the majority of considered articles (242 of 338) aimed to provide a wearable solution at the end. In contrast, 96 articles made use of a wearable device as a component of a more significant or complex solution.

\subsubsection{Does the work provide a prototype description}
Another question raised was whether the analyzed studies provide a device prototype description or not. This fact is particularly interesting, as papers providing only the general device description could be present.

The authors considered as a prototype any device with a minimum description presented, aside from being the ``main'' or ``one of the principal'' results produced by the paper.

\begin{figure}[ht]
  \centering
  \includegraphics[scale = 0.28]{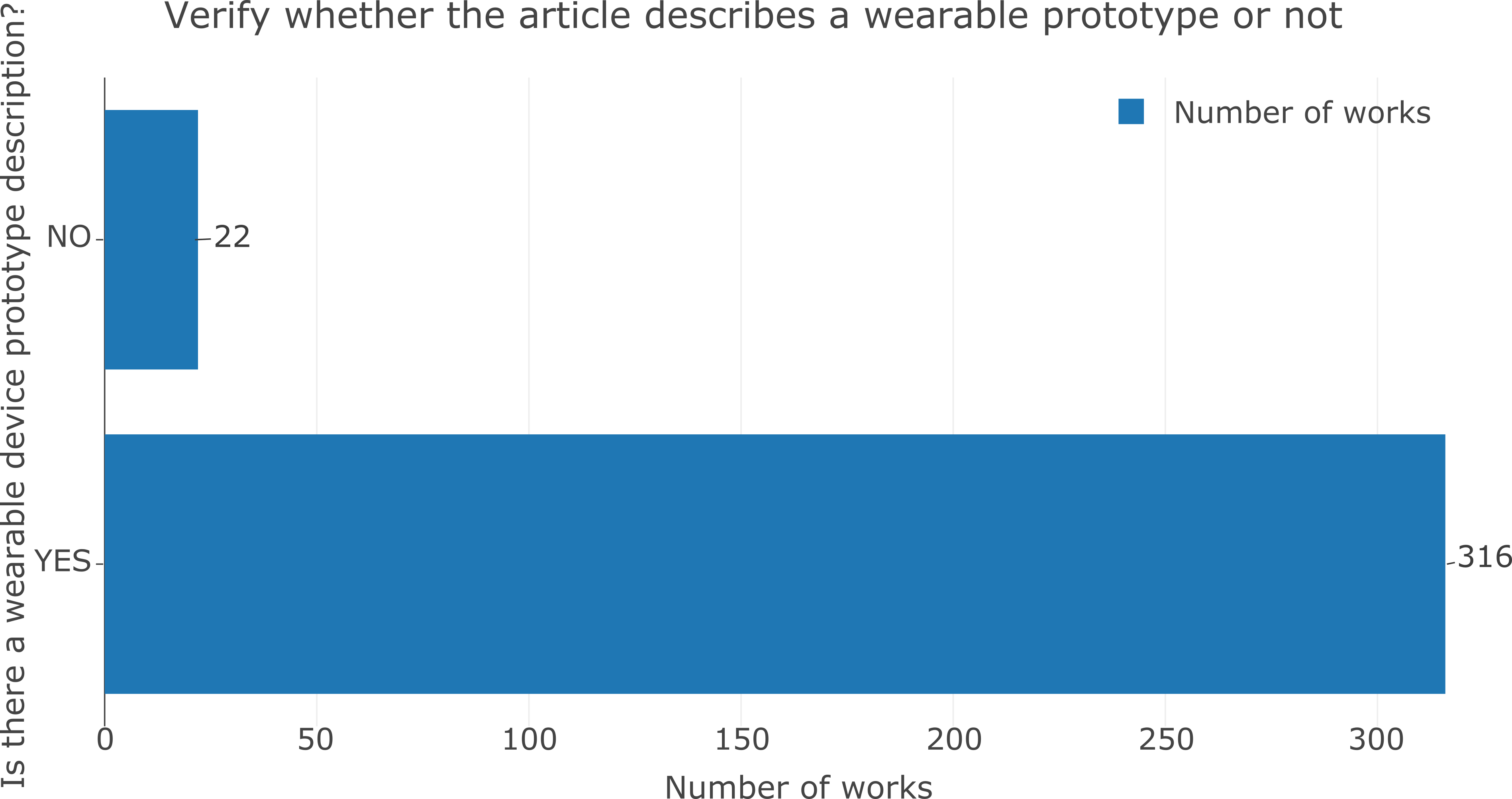}
  \caption{Verification of whether the analyzed studies present a wearable device prototype.}
  \label{fig:wearable_have_prot}
\end{figure}

As presented in Figure \ref{fig:wearable_have_prot}, 316 of the analyzed papers described a wearable device prototype with some level of detail. Conversely, only 22 papers did not present a final prototype description. As a general view of the wearable computing research context, this information infers that most scientists see the creation of real prototypes as a reasonable way to present their results.

\onecolumn 
\begin{center}
\begin{longtable}{ p{16.0em} | c | c | c | c | c }
\caption{Classification based on data extracted from the articles (2008 -- Apr/2017)}\\
\hline
\multicolumn{1}{c|}{\textbf{Reference}} & \multicolumn{1}{m{2.2em}|}{\textbf{Study type}} & \multicolumn{1}{m{6.5em}|}{\textbf{Work focuses on wearable devices?}} & \multicolumn{1}{m{6.5em}|}{\textbf{Prototype description?}} & \multicolumn{1}{m{6.5em}|}{\textbf{Wearable device description level}} & \multicolumn{1}{m{6.5em}}{\textbf{Wearable device complete solution?}} \\
\hline
\hline
\cite{Uddin:2015:WSF:2753509.2753513} \cite{7028031} \cite{7907202} \cite{4651159} & FW & -- & \checkmark & 2 & -- \\ 
\hline
\cite{Simon:2015:CSJ:2802083.2808401} \cite{7318575} \cite{7370275} & FW & \checkmark & \checkmark & 2 & -- \\  
\hline
\cite{Simon:2014:WJM:2634317.2634342} \cite{Nehani:2015:ENW:2753509.2753510} \cite{Nassani:2015:TAA:2818427.2818438} \cite{Nguyen:2016:IBR:2935643.2935649} \cite{Chan:2015:CWS:2702123.2702464} \cite{Wahl:2015:USE:2800835.2800914} \cite{Maeda:2016:WHA:2927929.2927946} \cite{Bailly:2012:SNP:2207676.2208576} \cite{Lane:2015:ZCD:2742647.2742672} \cite{Chien:2015:FMS:2800835.2800883} \cite{Dancu:2015:MNU:2785830.2785876} \cite{Mokaya:2015:MVB:2750858.2804258} \cite{Gartseev:2017:RPL:3045714.3045721} \cite{Munoz:2014:KEW:2559206.2580932} \cite{Hwang:2015:LYF:2702613.2732734} \cite{Scholl:2015:WWL:2750858.2807547} \cite{Webb:2016:WCG:2984511.2984564} \cite{Pfeiffer:2016:LYB:2935334.2935348} \cite{Takano:2014:ACA:2593968.2610455} \cite{Patankar:2016:WSO:2967878.2967893} \cite{Masai:2015:ATR:2800835.2800898} \cite{Pouryazdan:2016:WEP:2968219.2968286} \cite{6420443} \cite{7784835} \cite{4575008} \cite{7151345} \cite{7523697} \cite{6944272} \cite{6680610} \cite{6225000} \cite{7796223} \cite{4648016} \cite{7808003} \cite{7509332} \cite{7063093} \cite{6404520} \cite{7028005} \cite{5695932} \cite{6379440} \cite{6851442} \cite{7471436} \cite{7591799} \cite{Nogami2013} & FW & \checkmark & \checkmark & 3 & \checkmark \\ 
\hline
\cite{Liu:2015:TDU:2753509.2753518} & WP & \checkmark & \checkmark & 1 &  \checkmark\\ 
\hline
\cite{Perego:2015:UDW:2753509.2753522} \cite{Natarajan:2016:DAM:2971648.2971666} \cite{6575525} & FW & -- & \checkmark & 1 & -- \\ 
\hline
\cite{Chen:2015:EIE:2753509.2753517} & CS & -- & -- & 2 & \checkmark \\ 
\hline
\cite{Song:2015:HTE:2802083.2802092} \cite{Ha:2014:TWC:2594368.2594383} \cite{Kazemitabaar:2015:MTW:2771839.2771883} \cite{6427280} & VL & \checkmark & \checkmark & 4 & \checkmark \\ 
\hline
\cite{Yoon:2016:LUI:3016705.3016746} \cite{Schuldhaus:2016:YPM:2971763.2971772} \cite{Chu:2017:TWA:3024969.3025008} \cite{Chi:2015:WSC:2702123.2702451} \cite{Hamatani:2015:ECB:2695664.2695765} & VL & -- & -- & 0 & -- \\ 
\hline
\cite{Milosevic:2015:WIS:2753509.2753512} \cite{Natarajan:2013:DCU:2493432.2493496} & FW & -- & -- & 1 & \checkmark \\ 
\hline
\cite{Younes:2015:IAW:2800835.2801656} \cite{Sheng:2016:SAR:3013971.3014016} \cite{7738020} \cite{5109221} \cite{7591492} \cite{4601736} \cite{7379389} \cite{7545829} \cite{7008685} \cite{7080722} \cite{7319205} \cite{5955293} \cite{7474131} \cite{7810672} \cite{6607626} & FW & -- & \checkmark & 1 & \checkmark \\ 
\hline
\cite{Delabrida:2016:BWG:2903267.2903275} \cite{Popleteev:2015:ATI:2800835.2800938} \cite{Kalyanaraman:2015:ARC:2800835.2800856} \cite{Colley:2016:SHW:2875194.2875212} \cite{Antoniou:2015:PDI:2801948.2801992} \cite{Avila:2015:AWT:2786567.2794311} \cite{6746958} \cite{6726085} \cite{6038832} \cite{5229803} \cite{6696040} \cite{6610579} \cite{7394248} \cite{6637330} & FW & \checkmark & \checkmark & 1 & \checkmark \\ 
\hline
\cite{Roggen:2016:EFP:2971763.2971774} \cite{Ehleringer:2013:WLI:2468356.2468429} \cite{Kono:2016:WHS:2957265.2962652} \cite{Yataka:2011:CAP:1982185.1982274} \cite{Nguyen:2015:BAR:2829875.2829930} \cite{Ugulino:2015:LIW:2750858.2807541} \cite{Garcia:2015:WCI:2811780.2811959} \cite{Kritzler:2015:WTS:2836041.2836062} \cite{Zheng:2015:WSI:2702613.2725442} \cite{Pailes-Friedman:2015:BKA:2800835.2809435} \cite{Kosmalla:2016:CIP:2858036.2858562} \cite{Hiyama:2015:BPW:2800835.2800907} \cite{Twyman:2015:DWH:2677199.2680578} \cite{Scourboutakos:2017:PAR:3024969.3035534} \cite{Ranjan:2015:OHI:2750858.2804263} \cite{Song:2016:RWT:2851581.2890229} \cite{Zhang:2015:TWL:2807442.2807480} \cite{Ryokai:2014:EEH:2556288.2557225} \cite{Choi:2015:EWU:2832932.2856221} \cite{Brady:2015:CPB:2808006.2808039} \cite{Zhao:2016:EPW:2968219.2968577} \cite{Pfeiffer:2016:WFF:2851581.2890238} \cite{Peppler:2010:BLW:1810543.1810582} \cite{Voss:2016:SGD:2968219.2968310} \cite{Baldauf:2010:KWS:1785455.1785464} \cite{Zeagler:2016:SRD:2995257.2995390} \cite{Preuveneers:2016:PRH:2851613.2851683} \cite{Zhao:2016:MIS:2968120.2987726} \cite{Rantala:2015:HFG:2800835.2804334} \cite{Malu:2015:PWC:2702123.2702188} \cite{Fedosov:2016:DEW:3012709.3012721} \cite{Amores:2015:EDW:2702613.2732885} \cite{Song:2015:MWD:2702613.2725435} \cite{Guo:2015:AWC:2702613.2732755} \cite{Wang:2015:DMI:2800835.2800941} \cite{6379423} \cite{6778214} \cite{7416184} \cite{5342072} \cite{7155432} \cite{6151376} \cite{7015904} \cite{6974500} \cite{6461788} \cite{6420396} \cite{5702762} \cite{7422089} \cite{6375408} \cite{7460677} \cite{6812004} \cite{6610601} \cite{5652284} \cite{6563971} \cite{7759233} \cite{5627452} \cite{6778017} \cite{7177742} \cite{7451069} \cite{7427250} \cite{7430522} \cite{7533768} \cite{7133994} \cite{5447796} \cite{5204355} \cite{5571298} \cite{7344630} \cite{6420336} \cite{7749418} \cite{7463167} \cite{7110316} \cite{7270350} \cite{7894504} \cite{5627715} \cite{7888702} \cite{7899531} \cite{5759815} \cite{4649942} \cite{4911582} \cite{7592111} \cite{5090966} \cite{7484220} \cite{7303835} \cite{Majumder2017} \cite{Qaisar:2013:HMM:2600433.2600443} \cite{Cernea:2012:TWU:2404085.2404109} & FW & \checkmark & \checkmark & 2 & \checkmark \\ 
\hline
\cite{Ugulino:2015:PWS:2753509.2753515} & CS & \checkmark & \checkmark & 2 &  \checkmark\\ 
\hline
\cite{Chin:2015:DIA:2800835.2801667} \cite{Agarwal:2017:GWD:3041164.3041172} \cite{Sahni:2014:TEI:2634317.2634322} \cite{Roy:2015:FSB:2783449.2783520} \cite{Withana:2015:ZES:2702123.2702371} \cite{Grosse-Puppendahl:2015:ETS:2790044.2790059} \cite{Nirjon:2015:TWR:2742647.2742665} \cite{Kim:2016:HWH:2858036.2858196} \cite{Zaman:2014:KSL:2676431.2676433} \cite{Mokaya:2016:BWS:2959355.2959363} \cite{4487087} \cite{7393857} \cite{7098714} \cite{5229801} \cite{7177743} \cite{6987303} \cite{7319245} \cite{7059367} \cite{7888786} \cite{7236886} \cite{7294702} \cite{6878265} & FW & \checkmark & \checkmark & 4 & \checkmark \\ 
\hline
\cite{Maeda:2016:HWH:2988240.2988253} \cite{Klamet:2016:WWS:2948963.2948965} & WP & \checkmark & \checkmark & 2 & \checkmark \\ 
\hline
\cite{Bace:2017:HWA:3041164.3041203} \cite{Ng:2015:WCD:2786567.2794330} & CS & \checkmark & \checkmark & 2 & \checkmark  \\ 
\hline
\cite{Hamatani:2015:RCH:2753509.2753514} \cite{7868369} \cite{8a3a6a6c5bb444249c315f1651125165} & FW & -- & -- & 0 & -- \\ 
\hline
\cite{Ogawa:2014:EHF:2634317.2634337} & VL & -- & \checkmark & 2 & -- \\ 
\hline
\cite{Chan:2016:EWS:2992154.2996880} & IV & \checkmark & \checkmark & 2 & \checkmark \\ 
\hline
\cite{Badreddin:2015:TIP:2886444.2886482} \cite{Prieto:2016:TAT:2883851.2883927} & CS & -- & -- & 0 & -- \\ 
\hline
\cite{Muralidhar:2016:DWF:3012709.3012733} \cite{Seo:2017:IWE:3024969.3025057} \cite{Schneegass:2016:UOD:2914920.2915021} \cite{NunezPacheco:2017:TNS:3024969.3024979} \cite{Morrison:2016:ITA:2995257.2995391} \cite{Rajanna:2016:GGW:2857491.2857499} \cite{Washington:2016:WSI:2851581.2892282} \cite{Williams:2016:UWC:2968219.2972711} \cite{Dehzangi:2015:MBD:2750511.2750527} \cite{Jiang:2016:PDM:2968219.2971445} \cite{Mazilu:2015:WAG:2744352.2701431} \cite{Warren:2016:PSE:2839462.2856551} \cite{7299417} \cite{6423825} \cite{7840268} \cite{7289390} \cite{6611190} \cite{7518241} \cite{6548154} \cite{7844936} \cite{6903207} \cite{7727932} \cite{6346855} & VL & \checkmark & \checkmark & 2 & \checkmark  \\ 
\hline
\cite{Quintana:2016:KWE:2851581.2892493} & CS & -- & -- & 0 & \checkmark \\ 
\hline
\cite{Mokaya:2015:AMI:2800835.2801647} & WP & \checkmark & \checkmark & 4 & \checkmark \\ 
\hline
\cite{Jarusriboonchai:2015:CWP:2702613.2732833} \cite{Liu:2017:FCW:3024969.3025072} & WP & \checkmark & \checkmark & 3 & \checkmark \\ 
\hline
\cite{Hsieh:2015:TIG:2800835.2801661} \cite{4911577} \cite{6678519} \cite{7843664} \cite{7457112} \cite{7454554} \cite{7106937} \cite{5975448} \cite{6597819} \cite{6485186} \cite{7169792} \cite{7349409} \cite{7134078} \cite{7340953} \cite{4636308} \cite{5653821} \cite{6136953} \cite{7439470} \cite{7358841} \cite{6461277} \cite{6712463} \cite{6703792} & FW & -- & \checkmark & 2 & \checkmark \\ 
\hline
\cite{Lee:2016:DNW:2984511.2984583} \cite{Zhao:2015:HWL:2800835.2801670} \cite{Fang:2016:BER:2906388.2906411} \cite{Winoto:2015:SSE:2800835.2800915} \cite{Jylha:2015:WMI:2818346.2820763} \cite{Borsato:2016:EST:2857491.2857496} \cite{Haescher:2015:CCR:2769493.2769500} \cite{Vijay:2016:MMW:2839462.2856554} \cite{Knoop:2015:TCW:2702613.2732749} \cite{7015910} \cite{7127384} \cite{6107696} \cite{7452345} \cite{4571068} \cite{6688328} & VL & \checkmark & \checkmark & 3 & \checkmark \\  
\hline
\cite{Oertel:2016:MBS:2968219.2968574} & FW & -- & \checkmark & 0 & -- \\ 
\hline
\cite{Yang:2015:WAL:2742647.2742648} & FW & -- & \checkmark & 3 & -- \\ 
\hline
\cite{Gale:2015:HGP:2793107.2810309} & WP & \checkmark & -- & 0 & \checkmark \\ 
\hline
\cite{Mihoub:2017:SIM:3025171.3025195} \cite{Aromaa:2016:UWA:2994310.2994321} \cite{Cruz:2015:WMI:2836041.2836058} \cite{Albinali:2010:UWA:1864349.1864396} \cite{Lorincz:2009:MWS:1644038.1644057} \cite{Jee:2016:MPE:2993422.2993572} \cite{Altini:2013:CWA:2534088.2534106} \cite{7531452} \cite{7319248} \cite{5470653} & VL & -- & \checkmark & 2 & \checkmark \\ 
\hline
\cite{Ens:2015:CIR:2807442.2807449} & FW & -- & \checkmark & 0 & \checkmark \\ 
\hline
\cite{Ranjan:2016:TRP:2968219.2968279} \cite{Roinesalo:2016:SSD:2968219.2971350} & FW & \checkmark & \checkmark & 0 & \checkmark \\ 
\hline
\cite{Chen:2015:ULW:2757290.2757298} & VL & -- & \checkmark & 0 & -- \\ 
\hline
\cite{Magnusson:2016:IGW:2971485.2996752} & CS & \checkmark & \checkmark & 3 & \checkmark \\ 
\hline
\cite{Seifi:2016:EDS:2971763.2971783} & VL & \checkmark & \checkmark & 0 & \checkmark \\ 
\hline
\cite{Cumbo:2015:WAR:2838739.2838822} & VL & -- & -- & 0 & \checkmark \\ 
\hline
\cite{Nakajima:2016:IPB:3004010.3004041} \cite{Matsubara:2015:ERT:2800835.2804333} \cite{Liang:2016:PRC:2968219.2971462} \cite{7408495} \cite{7778023} \cite{5289412} \cite{7170674} & VL & -- & \checkmark & 1 & \checkmark \\ 
\hline
\cite{Thoring:2015:EDR:2702613.2732717} \cite{Hoang:2009:AIP:1862703.1862705} \cite{Spann:2017:ESL:3027385.3027427} & FW & -- & -- & 0 & \checkmark \\ 
\hline
\cite{Diep:2014:CBA:2676585.2676597} & VL & -- & -- & 2 & \checkmark \\ 
\hline
\cite{Rubin:2015:TMW:2750858.2805834} & IS & \checkmark & \checkmark & 1 & \checkmark \\ 
\hline
\cite{Kerz:2016:SWR:2968219.2971419} & CS & \checkmark & \checkmark & 1 & \checkmark \\ 
\hline
\cite{Goto:2015:WAS:2800835.2800863} \cite{Guo:2016:WSB:2971648.2971708} \cite{Haescher:2016:SUA:2851581.2892279} \cite{6271906} \cite{7836253} & VL & \checkmark & \checkmark & 1 & \checkmark \\ 
\hline
\cite{7319277} & VL & \checkmark & -- & 0 & \checkmark \\ 
\hline
\cite{7015908} & FW & \checkmark & -- & 0 & \checkmark  \\ 
\hline
\cite{7749528} \cite{6817874} \cite{6083808} \cite{7058423} \cite{7133581} & FW & -- & \checkmark & 3 & \checkmark \\  
\hline
\cite{7516233} \cite{6334932} & VL & \checkmark & \checkmark & 1 & --  \\ 
\hline
\cite{6094839} & IV & -- & \checkmark & 2 & \checkmark \\ 
\hline
\cite{5420163} & WP & -- & \checkmark & 2 & \checkmark \\ 
\hline
\cite{7866536} \cite{6775445} & IS & -- & \checkmark & 2 & \checkmark  \\ 
\hline
\cite{6290926} \cite{7463168} & FW & \checkmark & \checkmark & 4 & --  \\ 
\hline
\cite{5945506} & FW & -- & \checkmark & 4 & \checkmark  \\ 
\hline
\cite{7455975} & FW & \checkmark & \checkmark & 1 & -- \\ 
\hline
\cite{7230223} \cite{5415569} & FW & \checkmark & \checkmark & 3 & --  \\ 
\hline
\cite{6717926} & FW & \checkmark & \checkmark & 0 & --   \\ 
\hline
\cite{7556132} & IS & \checkmark & \checkmark & 4 & \checkmark  \\ 
\hline
\cite{ShingoSasaki2014CJ} & VL & -- & \checkmark & 1 & -- \\
\hline
\label{table:group1_results}
\end{longtable}
\end{center}
\twocolumn

\subsubsection{Wearable device description level} 
If there is a prototype described by the work, the question is how well this device is presented in the paper. Table \ref{table:extracted_data} shows five different degrees detailing the produced device according to how deep is its description:

\begin{itemize}
    \item \textbf{0}: Prototype is not described by the paper;
    \item \textbf{1}: Wearable device proposed by the article is lightly mentioned throughout the text;
    \item \textbf{2}: Prototype is mentioned in the text and demonstrated through the use of images;
    \item \textbf{3}: Device and its implementation process are described in the text and presented using images;
    \item \textbf{4}: Wearable device is deeply described by the text, with most (or all) of its hardware components being presented, as well as the design and build process.
\end{itemize}

\begin{figure}[ht]
  \centering
  \includegraphics[scale = 0.40]{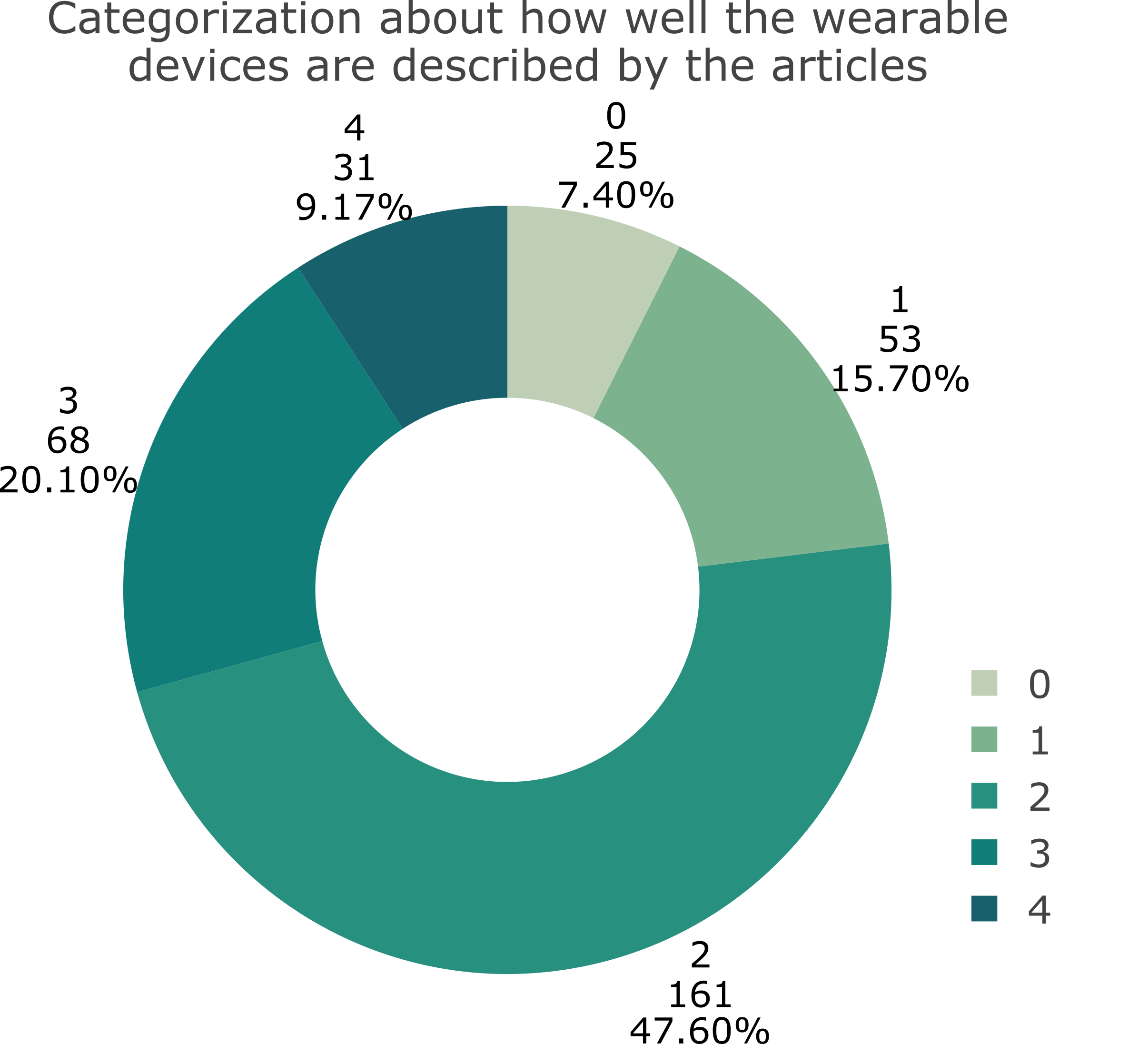}
  \caption{Wearable device prototypes description levels.}
  \label{fig:wearable_prot_desc_level}
\end{figure}

Analyzing the data presented in Figure \ref{fig:wearable_prot_desc_level}, it is possible to verify that the majority of papers provide a superficial description of the resulting device (Level 2: 161 or 47.60\%), while a more profound explanation (Level 4) is found in only 9.17\% of the articles. This result may suggest that scientists only describe the necessary parts so that readers can gain a basic understanding of their work, a fact also corroborated by the percentage of papers that do not provide a minimum description of the resulting device (Level 0: 25 or 7.40\%).

\subsubsection{Is the wearable prototype a complete solution}
The final question to be answered by this SRL concerns the scope of proposed wearable devices. Many articles include research that can be used in other research, or even by more general solutions involving embedded computing rather than wearable devices. Components such as sensors and vests/textiles technologies can apply to other solutions in different areas of research.

\begin{figure}[ht]
  \centering
  \includegraphics[scale = 0.28]{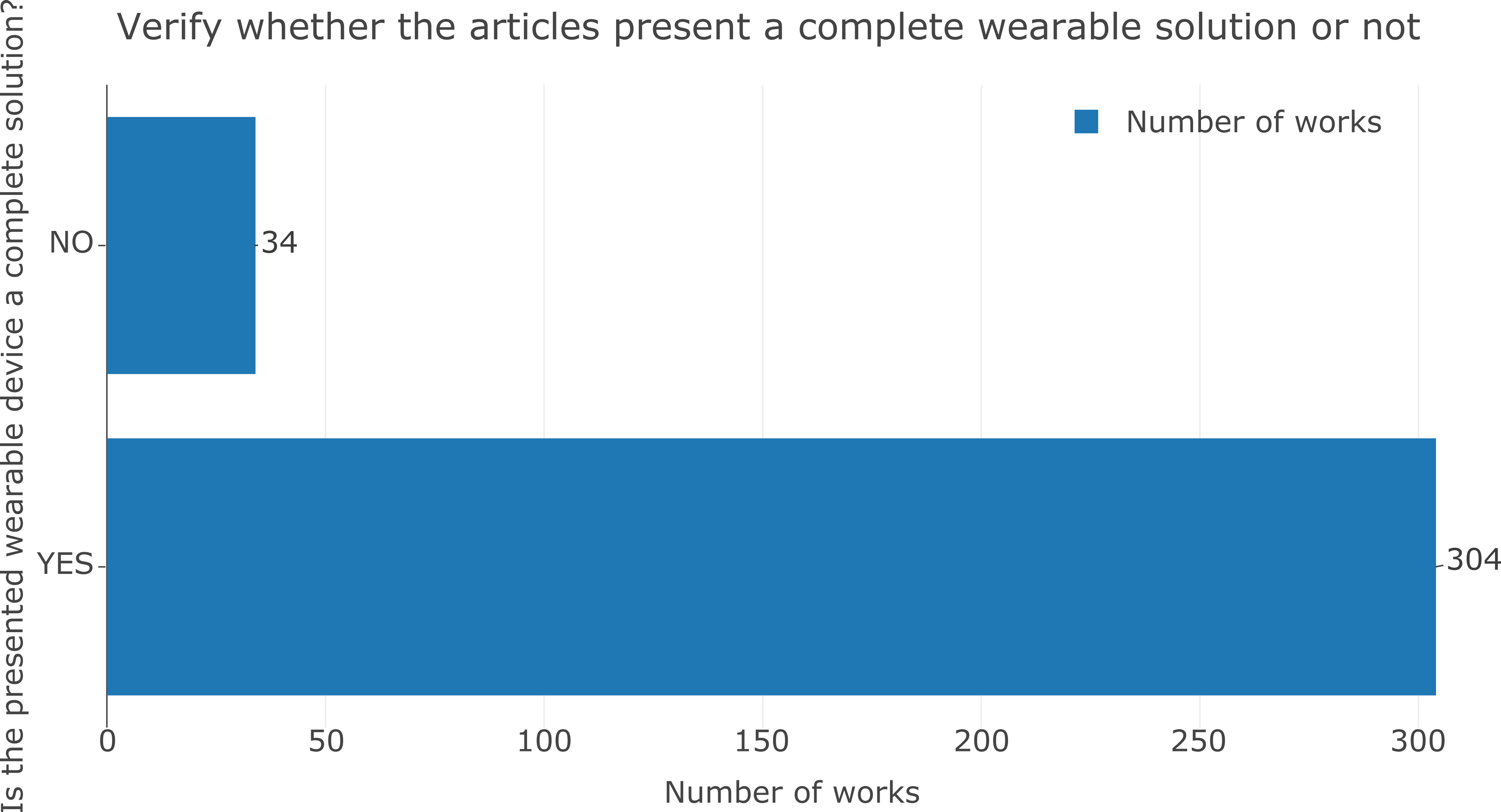}
  \caption{Verification of whether the considered prototypes are a complete wearable solution.}
  \label{fig:wearable_complete_solution}
\end{figure}

Data extracted from the papers can be used to create the chart presented in Figure \ref{fig:wearable_complete_solution}. Results show a massive number of works describing complete prototypes, usually contextualized inside an area trying to solve a specific or general problem, as demonstrated through the analysis in Section \ref{subsec:group_2}. 

The data presented here is also remarkable when taking into account the data already discussed in previous sections. Together this information can provide useful view regarding the nature of works being published in the wearable computing field. Through a superficial analysis, it is possible to affirm that scientists are focusing on complete prototypes, which commonly are presented at a reasonable completion level in their papers. 

The next section provides a more in-depth discussion of the main areas, problems, and general nature of studies within the context considered in this article.

\subsection{Group 2} \label{subsec:group_2}
The information presented here covers all the non-highlighted data listed in Table \ref{table:extracted_data}. Results presented for this group cover all articles (3,315). Except for the articles where data was manually extracted (338), all remaining papers were automatically processed by mining scripts. 

The results described here were also arranged according to the extracted data to which they refer. A set of specific charts and tables eases the task of comparing the items within each topic. Providing numerical values and visually comparable bars  to correlate results inside each topic display a weighted visual representation of the associated numerical values.

\subsubsection{General number of studies}
As previously described, studies involving a wearable device were screened from 2008 to 2017. Although this review did not completely cover 2017, it is possible to observe an increasing number of studies in this year. Increasing numbers of articles can be noted when grouping them by the publication year.

\begin{table}[!htbp]
\caption{Number of studies covering wearable devices (2008 -- Apr/2017)}
\begin{center}
\begin{tabular}{ c | c | c} 
\hline
\multicolumn{1}{c}{\textbf{Year}} & \multicolumn{1}{c}{\textbf{Number of studies}} & \multicolumn{1}{c}{\textbf{Variation from previous year}} \\ 
\hline
\hline
\rowcolor{Gray} 2008 & 171 & -- \\ 
2009 & 172 & +0,6\%\\ 
\rowcolor{Gray} 2010 & 192 & +11.6\%\\
2011 & 251 & +30,7\%\\ 
\rowcolor{Gray} 2012 & 235 & -6.4\%\\ 
2013 & 312 & +32,8\%\\ 
\rowcolor{Gray} 2014 & 422 & +35.2\%\\ 
2015 & 593 & +40,5\%\\ 
\rowcolor{Gray} 2016 & 772 & +30.2\%\\ 
2017/Apr & 195 & --\\ 
\hline
\textbf{Total/Max.} & \textbf{3315} & \textbf{+451,5\%}\\ 
\hline

\end{tabular}
\end{center}
\label{table:number_of_works_per_year}
\end{table}

\begin{figure}[!htbp]
  \centering
  \includegraphics[scale = 0.537]{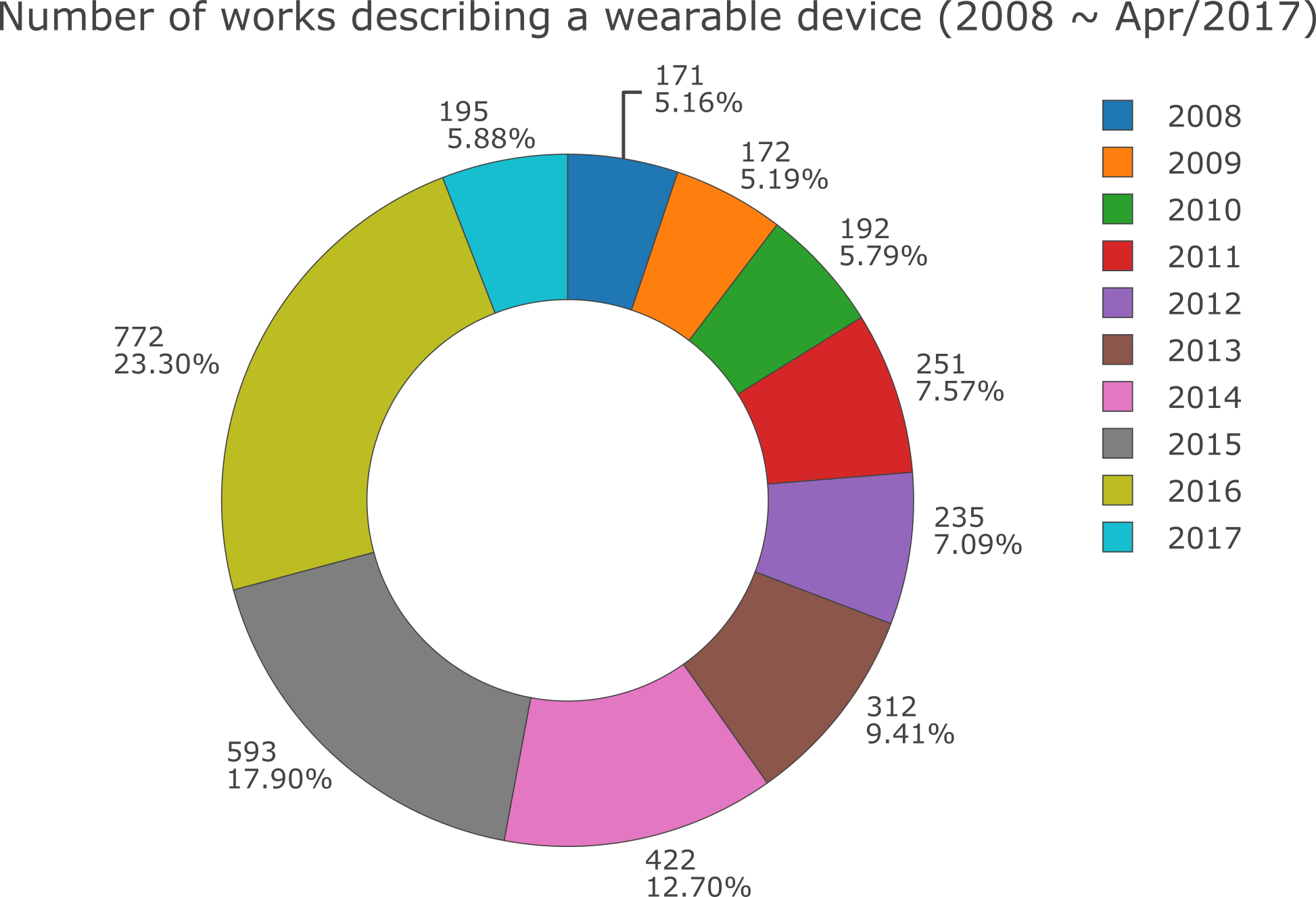}
  \caption{Percentage of wearables studies, separated by year.}
  \label{fig:wearables_devices_works}
\end{figure}

Figure \ref{fig:wearables_devices_works} presents a graphical separation and distribution of wearable devices studies over the past ten years, as considered by this review. It is possible to see that the number of studies in 2016 is more than four times the number in 2008. This reflects the increasing interest in wearables investigations, probably carried out in response to the demands generated by the market (and vice-versa). 

The number of published papers in 2017 corroborates this growth trend. In almost four months of analysis, more articles were published than during 2008, 2009, and 2010 combined. If this trend is maintained, 2017 will have almost as many (or even more) studies published than in 2016. 
 
The facts raised above highlight the importance of wearable devices in the research context. A more systematic view of this scenario can be found in Table \ref{table:number_of_works_per_year}, which shows the variation percentage year by year. Although the data regarding 2017 is not entirely presented, it tends to have numbers in the same range as 2016 if the same pace of first quarter was maintained for the remainder of 2017.

The real impact of wearables research can be seen by analyzing the last line of Table \ref{table:number_of_works_per_year}. It presents a value indicating an increase of at least 451.5\% in less than ten years (disregarding the last year).

\subsubsection{Affiliation country contribution}
Through the affiliation information, the authors have extracted data for associated countries. This type of information may be used to build a correlation between the interest level on wearables research and each country. Furthermore, as described in Section \ref{subsec:execution}, the filters implemented by this SRL allowed gathering of a set of more relevant articles, providing high-quality information.

\begin{table}[!htbp]
\caption{Wearable devices studies categorized by affiliation country (2008 -- Apr/2017)}
\begin{center}
\begin{tabular}{ c | l } 
\hline
\multicolumn{1}{c}{\textbf{Country}} & \multicolumn{1}{c}{\textbf{Contribution (\# of works)}}  \\ 
\hline
\hline
\rowcolor{Gray}USA & \makebox[0.5cm]{869} \mybar{3} \\ 
Japan & \makebox[0.5cm]{757} \mybar{2,613348677} \\ 
\rowcolor{Gray}China & \makebox[0.5cm]{344} \mybar{1,187571922} \\ 
Italy & \makebox[0.5cm]{292} \mybar{1,008055236} \\ 
\rowcolor{Gray}South Korea & \makebox[0.5cm]{246} \mybar{0,849252014}\\ 
United Kingdom & \makebox[0.5cm]{226} \mybar{0,780207135}\\ 
\rowcolor{Gray}Australia & \makebox[0.5cm]{224} \mybar{0,773302647}\\ 
Taiwan & \makebox[0.5cm]{164} \mybar{0,566168009}\\ 
\rowcolor{Gray}Canada & \makebox[0.5cm]{163} \mybar{0,562715765}\\ 
Germany & \makebox[0.5cm]{163} \mybar{0,562715765} \\ 
\rowcolor{Gray}Switzerland & \makebox[0.5cm]{130} \mybar{0,448791715}\\ 
India & \makebox[0.5cm]{121} \mybar{0,417721519}\\ 
\rowcolor{Gray}Spain & \makebox[0.5cm]{81} \mybar{0,279631761}\\ 
Hong Kong & \makebox[0.5cm]{69} \mybar{0,238204833}\\ 
\rowcolor{Gray}Singapore & \makebox[0.5cm]{61} \mybar{0,210586881}\\ 
Netherlands & \makebox[0.5cm]{53} \mybar{0,18296893}\\ 
\rowcolor{Gray}Portugal & \makebox[0.5cm]{49} \mybar{0,169159954}\\ 
Ireland & \makebox[0.5cm]{48} \mybar{0,16570771}\\ 
\rowcolor{Gray}France & \makebox[0.5cm]{45} \mybar{0,155350978}\\ 
Belgium & \makebox[0.5cm]{44} \mybar{0,151898734}\\ 
\rowcolor{Gray}Finland & \makebox[0.5cm]{38} \mybar{0,13118527}\\ 
Romania & \makebox[0.5cm]{34} \mybar{0,117376295}\\ 
\rowcolor{Gray}Sweden & \makebox[0.5cm]{34} \mybar{0,117376295}\\ 
Greece & \makebox[0.5cm]{33} \mybar{0,113924051}\\ 
\rowcolor{Gray}Malaysia & \makebox[0.5cm]{28} \mybar{0,096662831}\\ 
Brazil & \makebox[0.5cm]{23} \mybar{0,079401611}\\ 
\rowcolor{Gray}New Zealand & \makebox[0.5cm]{23} \mybar{0,079401611}\\ 
Slovenia & \makebox[0.5cm]{20} \mybar{0,069044879}\\ 
\rowcolor{Gray}Poland & \makebox[0.5cm]{19} \mybar{0,065592635}\\ 
Turkey & \makebox[0.5cm]{19} \mybar{0,065592635}\\ 
\rowcolor{Gray}Mexico & \makebox[0.5cm]{17} \mybar{0,058688147}\\ 
Thailand & \makebox[0.5cm]{16} \mybar{0,055235903}\\ 
\rowcolor{Gray}Denmark & \makebox[0.5cm]{15} \mybar{0,051783659}\\ 
Norway & \makebox[0.5cm]{13} \mybar{0,044879171}\\ 
\rowcolor{Gray}Israel & \makebox[0.5cm]{10} \mybar{0,03452244}\\ 
Austria & \makebox[0.5cm]{10} \mybar{0,03452244}\\ 
\hline

\end{tabular}
\end{center}
\label{table:number_of_works_per_country}
\end{table}

Table \ref{table:number_of_works_per_country} shows the countries that have been involved in at least ten (10) studies during the considered period (2008 -- Apr/2017). This information allows some conjecture regarding the relationship that has developed (or is under development) that countries have with studies on wearables. The initial places of this list are mainly composed of countries under these two statuses. Curiously, the sum of articles of the participation of the three first countries is higher than the sum of the next eleven (11) countries on the list. 

\begin{figure*}[!htbp]
  \centering
  \includegraphics[scale = 0.6]{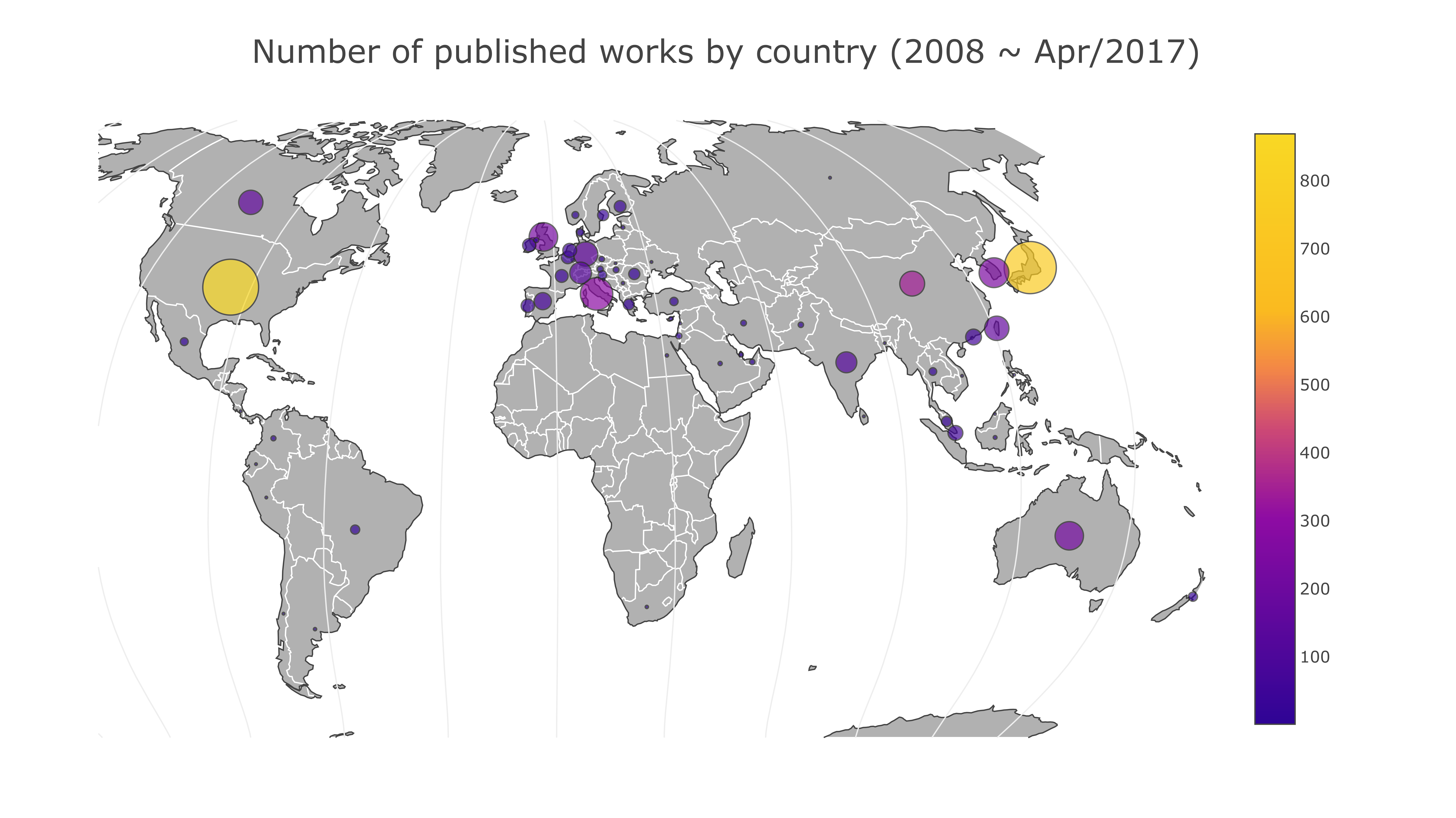}
  \caption{Number of published studies based on country.}
  \label{fig:wearables_devices_countries}
\end{figure*}

A graphical representation makes it possible to view the impact caused by the two first countries of the list in comparison with other participants. Figure \ref{fig:wearables_devices_countries} associates the number of works published by each country with their geographical position on the globe. The map shows an evident polarization of contributions by the countries of the northern hemisphere, particularly the European area and United States. However, relevant works can also be highlighted in the Oceania area (Australia, Indonesia, and so on) and in a few countries of South America.

In addition to the countries listed in this table, there are other nations with participation of less than ten works: Brunei, Slovakia, Philippines, Ecuador, Peru, Costa Rica, Ukraine, Lebanon, Chile, Russia, Bangladesh, Sri Lanka, Macau, Vietnam, Croatia, Egypt, South Africa, Serbia, Argentina, Indonesia, Luxembourg, Latvia, Cyprus, Saudi Arabia, Qatar, Colombia, Northern Ireland, Pakistan, Czech Republic, United Arab Emirates, Hungary, and Iran.

\subsubsection{Wearable type and on-body location} \label{subsubsec:device_location_type}
The extracted data makes it possible to categorize the articles according to the device type and on-body location where the user wears it. Initially, considering the analyzed articles, three major body areas were identified: Head, trunk, and limbs. This SRL associates every work to one of these three parts, highlighting the most common locations where devices are worn. In the second part of this subsection, a broader classification is presented, taking into account devices models and types.

\begin{figure*}[!htbp]
 \centering
 \includegraphics[scale = 0.7]{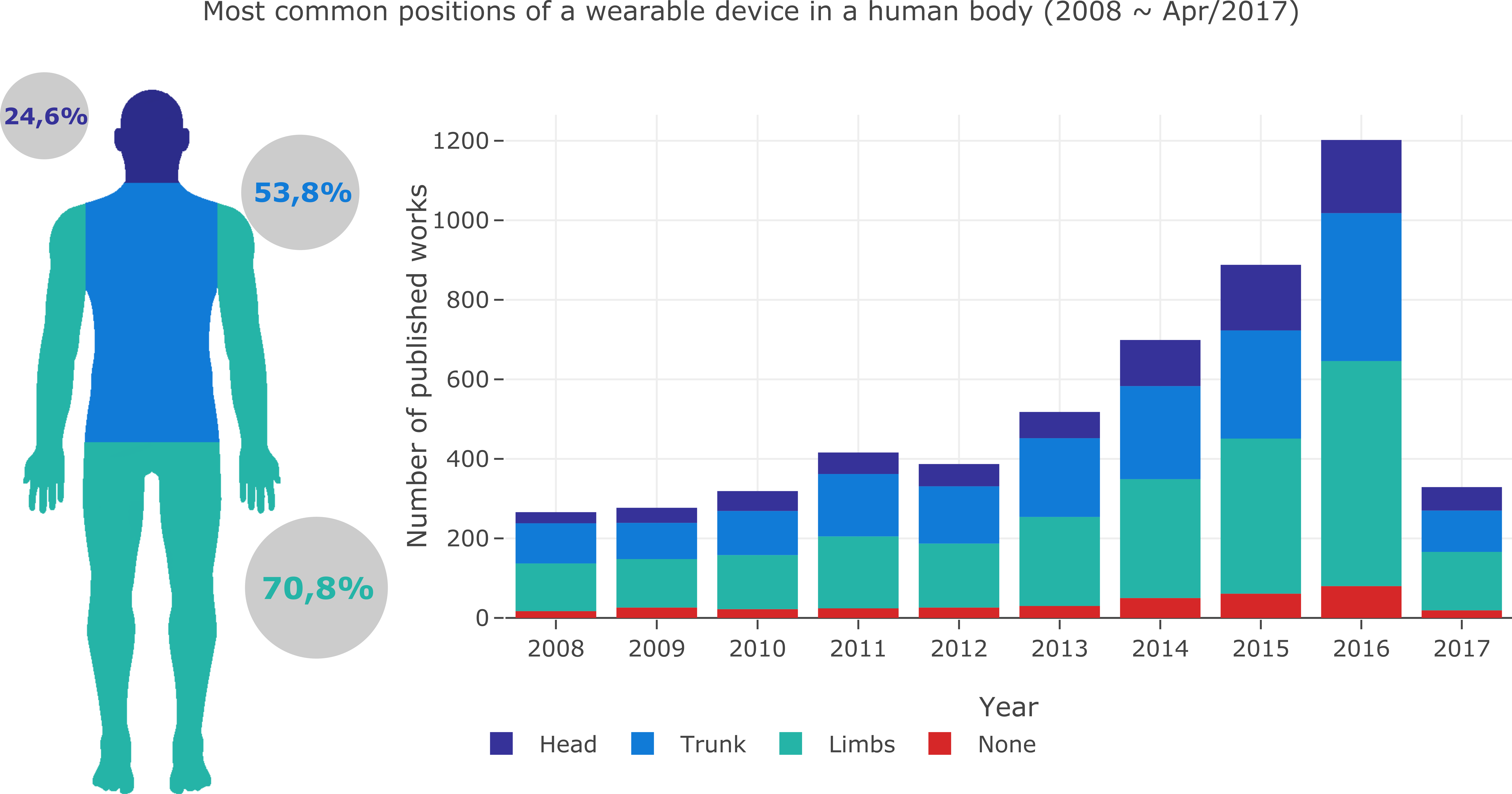}
 \caption{Distribution of wearable devices location on a human body.}
 \label{fig:wearable_type_infographic}
\end{figure*}

Figure \ref{fig:wearable_type_infographic} summarizes the most common locations where a wearable device is attached to a human body. Data presented by this figure consider the location where the user can wear a solution. Note that the sum of percentages will be higher than 100\%, as the same device/solution can be worn in different body locations. Resulting data indicates that the ``Limbs'' (70.8\%) are the most common location to attach a wearable, followed by the ``Trunk'' (53.8\%) and finally the ``Head'' (24.6\%). The chart on the right shows how the devices’ wearability has evolved over the last ten years. From 2014 onward, there is a relevant discrepancy regarding the number of wearables developed to be worn on ``Limbs'' in comparison to the other two regions. It is also valid to highlight that, in every year there were a relevant number of works that did not describe where the device possibly fits on the body. As in previous subsections, this chart also shows a similar curve of evolution following the number of articles published in the considered period.

Given the most common places where a wearable can be attached, a more detailed investigation can characterize the type of device and more specific places where devices are worn. Table \ref{table:wearable_type} shows a relationship between the wearable type/``location place'' with the number of studies in which they appear. Data presented in the second column counts the number of articles that index devices of the type listed by the first column. This type of characterization is particularly difficult, as there is no general classification pattern applicable to all studies, resulting in some discrepancies and redundancies. This is the main reason why the sum of all studies exceeds the total number of papers analyzed by this SRL. However, these repetitions cannot be ignored because they present essential patterns and behaviors. 

Initially, body locations with a high number of referring articles are the general ones: ``Arm-worn'', ``Wrist-worn'', ``Chest-worn'', and so on. This scenario occurs because some papers do not precisely specify the location where a device can be worn. Moreover, the two first most-cited positions (``Arm-worn'' and ``Wrist-worn'') suggest a significant use of devices such as watches, bands, and wrist strips. Corroborating the data presented in Figure \ref{fig:wearable_type_infographic}, the latest positions in Table \ref{table:wearable_type} are filled with solutions worn in the head region, such as ``In-ear'', ``Hat'', ``Headphone'', and so on.

\begin{table}[!htbp]
\caption{Most common types of wearable devices (2008 -- Apr/2017)}
\begin{center}
\begin{tabular}{ c | l } 
\hline
\multicolumn{1}{c}{\textbf{Wearable device type/location}} & \multicolumn{1}{c}{\textbf{Number of works}}  \\ 
\hline
\hline
\rowcolor{Gray}Arm-worn & \makebox[0.5cm]{682} \mybar{3}\\ 
Wrist-worn & \makebox[0.5cm]{642} \mybar{2,824046921}\\ 
\rowcolor{Gray}Chest-worn & \makebox[0.5cm]{580} \mybar{2,551319648}\\ 
Finger-worn & \makebox[0.5cm]{475} \mybar{2,089442815}\\ 
\rowcolor{Gray}Leg-worn & \makebox[0.5cm]{466} \mybar{2,049853372}\\ 
Knee-worn & \makebox[0.5cm]{325} \mybar{1,429618768}\\ 
\rowcolor{Gray}Glass & \makebox[0.5cm]{322} \mybar{1,416422287}\\ 
Belt & \makebox[0.5cm]{298} \mybar{1,31085044}\\ 
\rowcolor{Gray}Ankle-worn & \makebox[0.5cm]{277} \mybar{1,218475073}\\ 
Waist-worn & \makebox[0.5cm]{260} \mybar{1,143695015}\\ 
\rowcolor{Gray}Shoulder-worn & \makebox[0.5cm]{243} \mybar{1,068914956}\\ 
Thigh-worn & \makebox[0.5cm]{183} \mybar{0,804985337}\\ 
\rowcolor{Gray}Shirt & \makebox[0.5cm]{182} \mybar{0,80058651}\\ 
Vest & \makebox[0.5cm]{176} \mybar{0,774193548}\\ 
\rowcolor{Gray}Shoes sensors & \makebox[0.5cm]{166} \mybar{0,730205279}\\ 
Neck-worn & \makebox[0.5cm]{164} \mybar{0,721407625}\\ 
\rowcolor{Gray}Wristband & \makebox[0.5cm]{158} \mybar{0,695014663}\\ 
Exoskeleton & \makebox[0.5cm]{153} \mybar{0,673020528}\\ 
\rowcolor{Gray}Glove & \makebox[0.5cm]{150} \mybar{0,659824047}\\ 
Patch & \makebox[0.5cm]{150} \mybar{0,659824047}\\ 
\rowcolor{Gray}SmartWatch & \makebox[0.5cm]{127} \mybar{0,558651026}\\ 
Bag & \makebox[0.5cm]{72} \mybar{0,316715543}\\ 
\rowcolor{Gray}Helmet & \makebox[0.5cm]{71} \mybar{0,312316716}\\ 
Strip & \makebox[0.5cm]{66} \mybar{0,290322581}\\ 
\rowcolor{Gray}Backpack & \makebox[0.5cm]{63} \mybar{0,2771261}\\ 
In-ear & \makebox[0.5cm]{62} \mybar{0,272727273}\\ 
\rowcolor{Gray}Hat & \makebox[0.5cm]{56} \mybar{0,246334311}\\ 
Headphone & \makebox[0.5cm]{53} \mybar{0,23313783}\\ 
\rowcolor{Gray}Necklace & \makebox[0.5cm]{44} \mybar{0,193548387}\\ 
Headband & \makebox[0.5cm]{33} \mybar{0,14516129}\\ 
\rowcolor{Gray}Goggles & \makebox[0.5cm]{27} \mybar{0,118768328}\\ 
Backbone & \makebox[0.5cm]{18} \mybar{0,079178886}\\ 
\rowcolor{Gray}SmartBand & \makebox[0.5cm]{17} \mybar{0,074780059}\\ 
Head-worn & \makebox[0.5cm]{16} \mybar{0,070381232}\\ 
\rowcolor{Gray}Jewel & \makebox[0.5cm]{7} \mybar{0,030791789}\\ 
Halo & \makebox[0.5cm]{3} \mybar{0,013196481}\\ 
\hline
\end{tabular}
\end{center}
\label{table:wearable_type}
\end{table}

\begin{figure*}[!htbp]
 \centering
 \includegraphics[scale = 0.6]{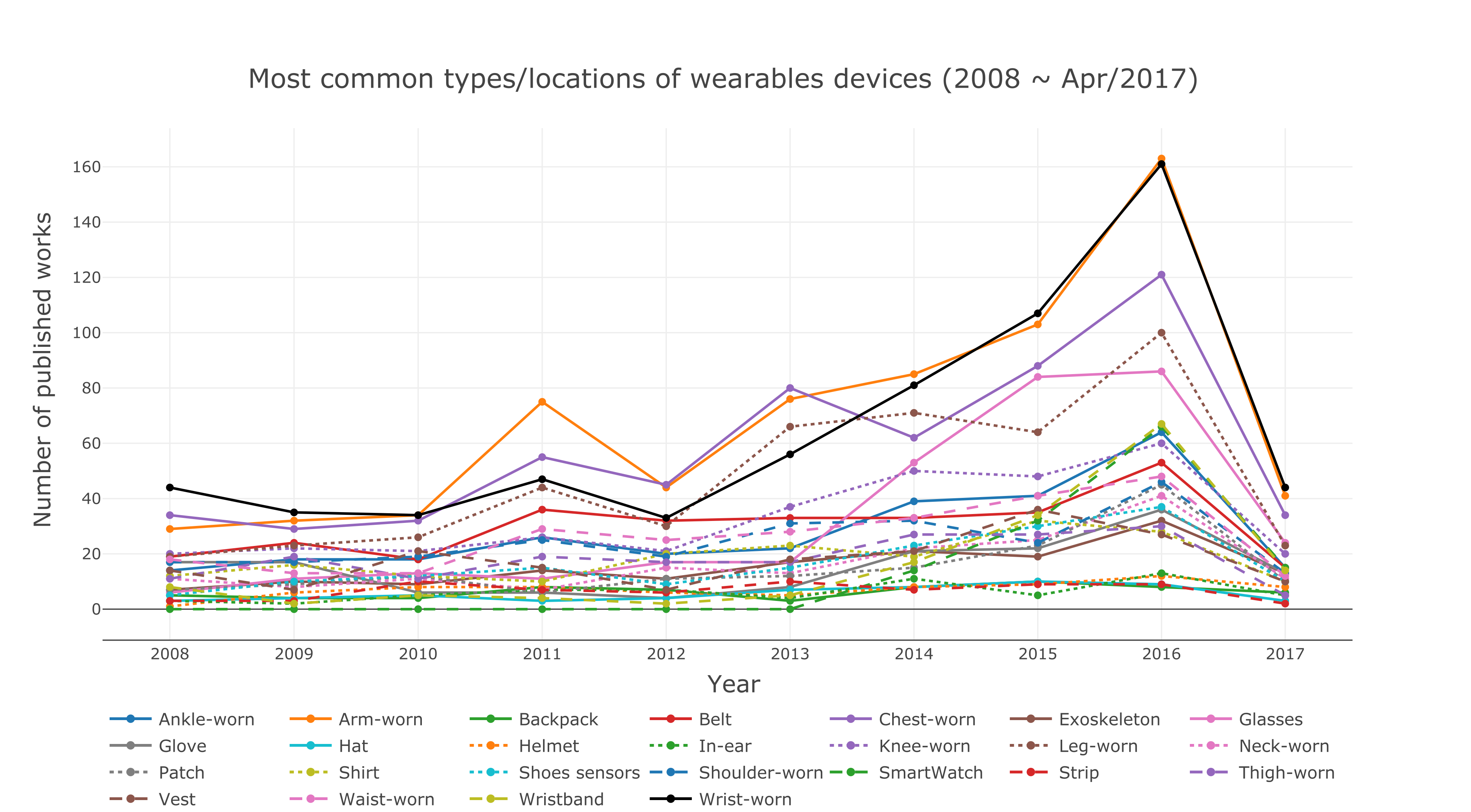}
 \caption{Most common device types and locations to attach a wearable device according to the research published during the considered period.}
 \label{fig:wearable_type_details}
\end{figure*}

A trending analysis of data as presented by Table \ref{table:wearable_type} is shown in Figure \ref{fig:wearable_type_details} through the information discretization considering the past years. The related chart yields a set of some interesting conclusions:
\begin{itemize}
    \item ``Arm-worn'' and ``Wrist-worn'' research solutions are currently the most popular ones. However, between 2011–2013 there was an increasing interest in devices attached to the user’s chest (``Chest-worn'');
    \item Devices attached to the user’s leg (``Leg-worn'') were not popular in 2008. On the other hand, this type of device indicates growing interest in research in recent years, reaching the fourth ranking in 2016 and 2017;
    \item A more incisive evolution can be observed when taking ``Glasses'' into account. From a reduced number of references in 2008, these devices have significantly evolved during later years. A possible reason for this behavior may be associated with the ``boom'' of Augmented Reality (AR) and Virtual Reality (VR) devices that began in 2012. This is confirmed by the data presented in Figure \ref{fig:wearables_devices_technologies}.   
\end{itemize}  

As with previous results, the chart described in Figure \ref{fig:wearable_type_details} also presents the same curve tendency, reflecting the increasing number of studies conducted during the latest years.

\subsubsection{Applicability areas}\label{subsubsec:applicability_areas}
Subsection \ref{subsubsec:applicability} has presented the applicability relevance of various wearable devices. This subsection shows the results regarding information extracted by considering applicability. This type of data is particularly important to indicate past, current, and future trends.      

\begin{table}[!htbp]
\caption{Wearable devices studies categorized by applicability area (2008 -- Apr/2017)}
\begin{center}
\begin{tabular}{ c | l } 
\hline
\multicolumn{1}{c}{\textbf{Applicability Area}} & \multicolumn{1}{c}{\textbf{Number of works}}  \\ 
\hline
\hline
\rowcolor{Gray}Healthcare/medicine & \makebox[0.5cm]{1733} \mybar{3} \\  
Environm./objects and ind. sensing & \makebox[0.5cm]{1284} \mybar{2,222735141} \\ 
\rowcolor{Gray}Sports & \makebox[0.5cm]{397} \mybar{0,687247548} \\ 
Entertainment & \makebox[0.5cm]{265} \mybar{0,458742066}\\ 
\rowcolor{Gray}Security/Rescue & \makebox[0.5cm]{243} \mybar{0,420657819}\\ 
Textile & \makebox[0.5cm]{236} \mybar{0,408540104}\\ 
\rowcolor{Gray}Navigation & \makebox[0.5cm]{235} \mybar{0,406809002}\\ 
Fabric & \makebox[0.5cm]{234} \mybar{0,4050779}\\ 
\rowcolor{Gray}Education & \makebox[0.5cm]{206} \mybar{0,35660704}\\ 
Civil construction & \makebox[0.5cm]{190} \mybar{0,328909406}\\ 
\rowcolor{Gray}Artistic & \makebox[0.5cm]{152} \mybar{0,263127525}\\ 
Maintenance & \makebox[0.5cm]{98} \mybar{0,169648009}\\ 
\rowcolor{Gray}Military & \makebox[0.5cm]{78} \mybar{0,135025967}\\ 
Learning/Teaching & \makebox[0.5cm]{68} \mybar{0,117714945}\\ 
\rowcolor{Gray}Cooking & \makebox[0.5cm]{38} \mybar{0,065781881}\\ 
Agriculture & \makebox[0.5cm]{26} \mybar{0,045008656}\\ 
\rowcolor{Gray}3D modeling & \makebox[0.5cm]{11} \mybar{0,019042123} \\
Elect. devices interaction & \makebox[0.5cm]{6} \mybar{0,010386613}\\ 
\rowcolor{Gray}Geology & \makebox[0.5cm]{2} \mybar{0,003462204}\\ 
\hline
\end{tabular}
\end{center}
\label{table:number_of_works_per_applicability_area}
\end{table}    

Data presented in this subsection was grouped considering only the most common applicability areas. It is also valid to note that the same published article may appear in two or more applicability areas owing to its scope. Table \ref{table:number_of_works_per_applicability_area} presents an overview of applicability area popularity within the considered window of time. Through a deeper analysis, it is possible to highlight the wearables importance in healthcare and medicine, as they encompass most of the studies. Because of its nature, the second most addressed area of applicability (``Environm./individual and object sensing'') may be used in healthcare/medicine wearables, as it commonly overlaps human body sensing for healthcare purposes. In addition, another essential index can be obtained when comparing the sum of the studies inside the two first applicability areas, which is higher than the sum of all remaining areas.

\begin{figure*}[!htbp]
  \centering
  \includegraphics[scale = 0.57]{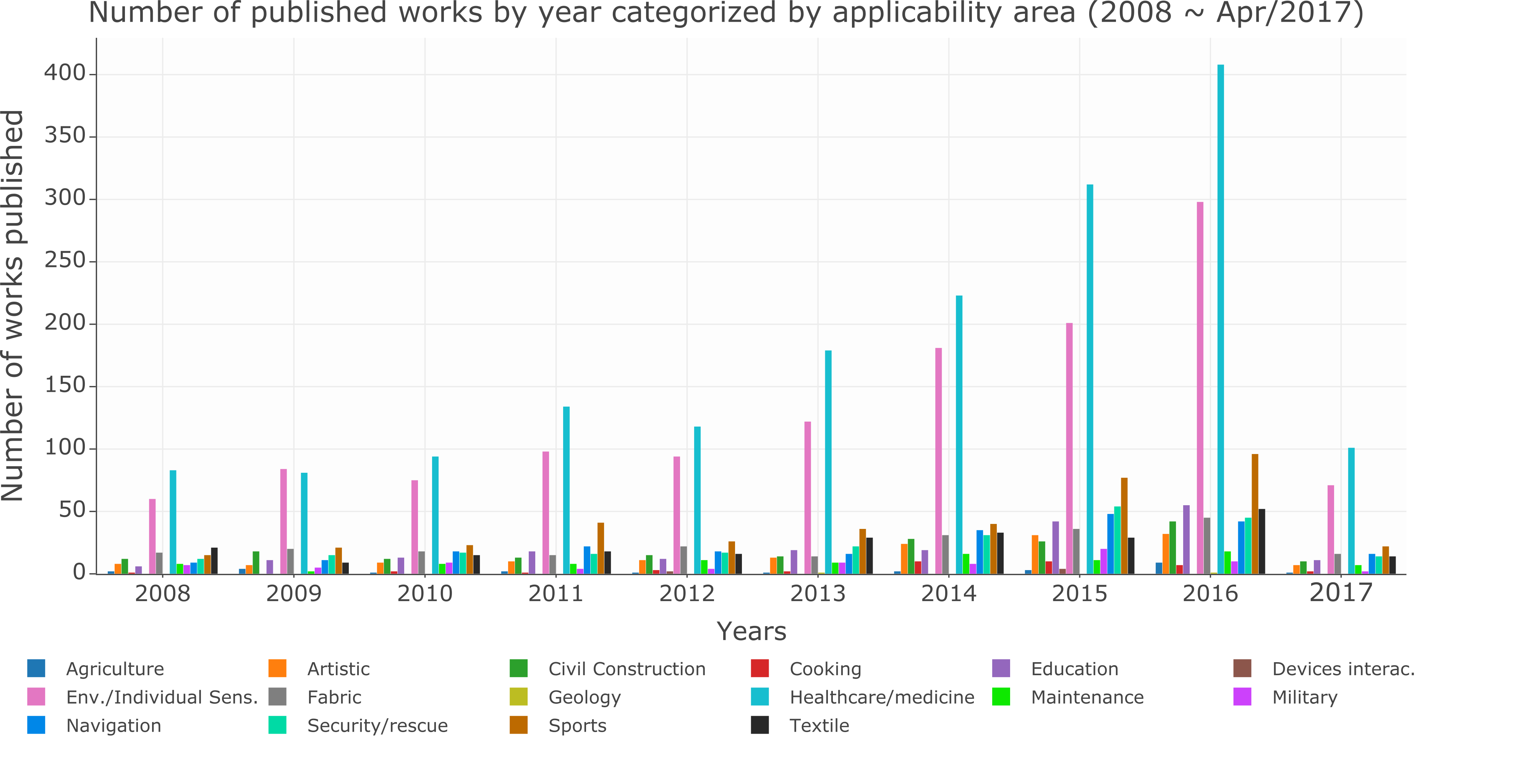}
  \caption{Number of published studies separated into applicability areas during considered years.}
  \label{fig:wearables_devices_applicability}
\end{figure*}

A graphical analysis regarding the impact of each applicability area is shown in Figure \ref{fig:wearables_devices_applicability}. Despite the most representative areas, this figure also shows the trends in wearable devices research during the last ten years. As previously concluded from Table \ref{table:number_of_works_per_applicability_area}, here it is also possible to verify the increasing number of papers addressing ``Healthcare/medicine'' and ``Environm./individual and object sensing''. Additionally, the ``Sports'' area should also receive attention, as this topic has been involved in a significant number of studies during recent times. Conversely, hot topics at the end of last decade have declined in interest, such as ``Textile'' and ``Fabric''. Mention of ``Education'', ``Security/rescue'', and ``Navigation'' areas is warranted, as they have maintained the same growth rhythm over the years.  

Finally, when analyzing just 2017, it is fairly obvious that the same tendencies described above are maintained, although in a reduced number, as just the beginning of the year was included in the SRL.

\subsubsection{Research focus} \label{subsubsec:addressed_problems}
As crucial as the applicability areas are for the problems addressed by the wearable devices, this subsection presents the most common research focus. As a means to reduce the analysis scope, only the most frequent problems were listed. Presented data can be employed to estimate hot-trending subjects, as well as to change the focus of scientific investigations. As reported previously, the research focus list of considered topics was manually extracted from a set of 338 articles used as samples. The remaining papers were then automatically analyzed using these 338 papers as a baseline.

\begin{table}[!htbp]
\caption{Wearable devices studies categorized according to research focus (2008 -- Apr/2017)}
\begin{center}
\begin{tabular}{ c | l } 
\hline
\multicolumn{1}{c}{\textbf{Research focus}} & \multicolumn{1}{c}{\textbf{Number of works}}  \\ 
\hline
\hline
\rowcolor{Gray}Gait assistance/support/tracking & \makebox[0.5cm]{965} \mybar{3}\\ 
Diseases/Disturbances detection & \makebox[0.5cm]{789} \mybar{2,452849741}\\ 
\rowcolor{Gray}Posture and gesture recognition & \makebox[0.5cm]{668} \mybar{2,076683938}\\ 
Rehabilitation & \makebox[0.5cm]{450} \mybar{1,398963731}\\ 
\rowcolor{Gray}Human act. / Act. of daily living & \makebox[0.5cm]{338} \mybar{1,050777202}\\ 
Stress management/monitoring & \makebox[0.5cm]{311} \mybar{0,966839378}\\ 
\rowcolor{Gray}Fall detection & \makebox[0.5cm]{270} \mybar{0,839378238}\\ 
Sleep staging/monitoring & \makebox[0.5cm]{261} \mybar{0,811398964}\\ 
\rowcolor{Gray}Daily life monitoring & \makebox[0.5cm]{253} \mybar{0,786528497}\\ 
User interface/user experience & \makebox[0.5cm]{251} \mybar{0,780310881}\\ 
\rowcolor{Gray}Breath analysis & \makebox[0.5cm]{226} \mybar{0,702590674}\\ 
Cognitive assistance & \makebox[0.5cm]{171} \mybar{0,531606218}\\ 
\rowcolor{Gray}Energy efficiency & \makebox[0.5cm]{162} \mybar{0,503626943}\\ 
Privacy & \makebox[0.5cm]{125} \mybar{0,388601036}\\ 
\rowcolor{Gray}Visually impaired support & \makebox[0.5cm]{123} \mybar{0,38238342}\\ 
Remote monitoring & \makebox[0.5cm]{89} \mybar{0,276683938}\\ 
\rowcolor{Gray}Context awareness & \makebox[0.5cm]{78} \mybar{0,242487047}\\ 
Personal energy expenditure & \makebox[0.5cm]{78} \mybar{0,242487047}\\ 
\rowcolor{Gray}Physiological parameters analysis & \makebox[0.5cm]{76} \mybar{0,23626943}\\ 
Gaze tracking & \makebox[0.5cm]{65} \mybar{0,202072539}\\ 
\rowcolor{Gray}Eye tracking & \makebox[0.5cm]{43} \mybar{0,133678756}\\ 
Dementia support/monit./therapy & \makebox[0.5cm]{39} \mybar{0,121243523}\\ 
\rowcolor{Gray}Indoor/Outdoor localiz./navig. & \makebox[0.5cm]{35} \mybar{0,10880829}\\ 
Autism therapy/support & \makebox[0.5cm]{26} \mybar{0,080829016}\\ 
\rowcolor{Gray}Brain-related diseases & \makebox[0.5cm]{13} \mybar{0,040414508}\\ 
Dead reckoning navigation & \makebox[0.5cm]{12} \mybar{0,037305699}\\ 
\hline
\end{tabular}
\end{center}
\label{table:number_of_works_per_problem}
\end{table}

As listed in subsection \ref{subsubsec:problems}, problems considered by this SRL are the most common ones frequently addressed by wearable research. Table \ref{table:number_of_works_per_problem} presents articles categorized and indexed according to the problem they discuss. The second column shows the number of studies in which each topic was listed. This data provides a representative insight regarding hot-topics/problems currently discussed in the research literature. Reflecting the applicability analysis (subsection \ref{subsubsec:applicability_areas}), the four most addressed problems are associated with healthcare/medicine applicability areas. Some of the remaining topics can also be contextualized inside this area, although this association may not be directly presented or discussed in any of the considered articles.

\begin{figure*}[!htbp]
  \centering
  \includegraphics[scale = 0.6]{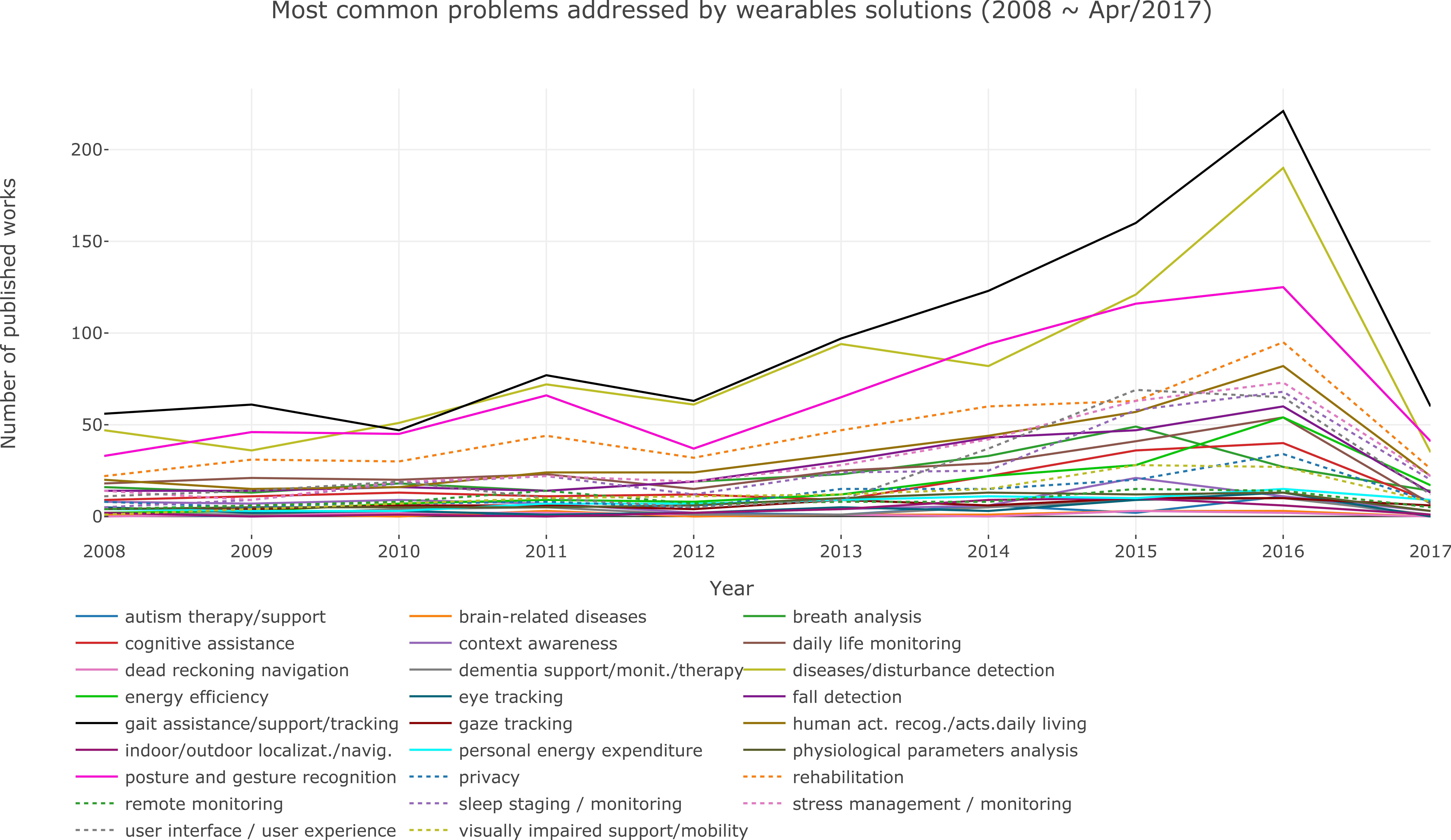}
  \caption{Most common problems addressed by wearables research during considered period.}
  \label{fig:wearables_devices_problems}
\end{figure*}

More relevant and in-depth analysis can be performed through Figure \ref{fig:wearables_devices_problems}. This figure shows how the research on and interest in each problem have evolved, year by year. Although the ``Gait assistance/support/tracking'' problem is considered by a large number of works, there is also a relevant number of articles covering topics such as the detection of diseases or general disturbances (``Diseases/disturbances detection''). From this figure, initially less relevant problems have gained prominence over the years, such as the sleep quality monitoring (``Sleep stating/monitoring'') and wearables applied to general rehabilitation. 

Finally, in the bottom part of Figure \ref{fig:wearables_devices_problems}, it is possible to observe a significant number of less-addressed subjects. Knowing the most relevant areas may allow scientists to change their research focus on current or future work. Additionally, although the fact that the three most addressed topics have not changed over the years, this review has considered only the relevant problems, with a set of minor problems not being listed in Table \ref{table:number_of_works_per_problem}.  

\subsubsection{Technologies, methods, and techniques} \label{subsubsec:common_technologies_methods}
This type of information allows verification of the most trending topics based on a set of different options, providing a snapshot of trending areas in wearable devices. As with previous analyses, data subjects presented by this subsection were initially manually extracted from 338 papers randomly selected, while the remaining articles underwent an automatic mining script. To summarize the retrieved data, only most relevant technologies, methods, and techniques are described below.

\begin{table}[!htbp]
\caption{Wearable devices studies categorized by the technologies, methods, and techniques they address (2008 -- Apr/2017)}
\begin{center}
\begin{tabular}{ c | l } 
\hline
\multicolumn{1}{c}{\textbf{Tech./method or technique}} & \multicolumn{1}{c}{\textbf{Number of works}}  \\ 
\hline
\hline
\rowcolor{Gray}Artificial Intelligence & \makebox[0.5cm]{870} \mybar{3}\\ 
Haptic & \makebox[0.5cm]{380} \mybar{1,310344828}\\ 
\rowcolor{Gray}AR/VR & \makebox[0.5cm]{330} \mybar{1,137931034}\\ 
Root Mean Square & \makebox[0.5cm]{289} \mybar{0,996551724}\\ 
\rowcolor{Gray}Heart Rate Variability (HRV) & \makebox[0.5cm]{168} \mybar{0,579310345}\\ 
Phtopletysmography (PPG) & \makebox[0.5cm]{158} \mybar{0,544827586}\\ 
\rowcolor{Gray}Fast Fourier Transform & \makebox[0.5cm]{147} \mybar{0,506896552}\\ 
HMD/HUD & \makebox[0.5cm]{108} \mybar{0,372413793} \\
\rowcolor{Gray}Internet of Things (IoT) & \makebox[0.5cm]{105} \mybar{0,362068966}\\ 
Brain-computer interf. (BCI) & \makebox[0.5cm]{60} \mybar{0,206896552}\\ 
\rowcolor{Gray}Beacons & \makebox[0.5cm]{52} \mybar{0,179310345}\\ 
Object recognition & \makebox[0.5cm]{52} \mybar{0,179310345}\\ 
\rowcolor{Gray}Dynamic Time Warping & \makebox[0.5cm]{50} \mybar{0,172413793}\\ 
Graphene & \makebox[0.5cm]{42} \mybar{0,144827586}\\ 
\rowcolor{Gray}Electrodermal Activity (EDA) & \makebox[0.5cm]{35} \mybar{0,120689655}\\ 
Face recognition & \makebox[0.5cm]{27} \mybar{0,093103448}\\ 
\rowcolor{Gray}Least Mean Square (LMS) & \makebox[0.5cm]{27} \mybar{0,093103448}\\ 
Solar Energy & \makebox[0.5cm]{25} \mybar{0,086206897}\\ 
\rowcolor{Gray}Deep Brain Stimulation (DBS) & \makebox[0.5cm]{17} \mybar{0,05862069}\\ 
Electrical Imp. Tomography (EIT) & \makebox[0.5cm]{17} \mybar{0,05862069}\\ 
\hline
\end{tabular}
\end{center}
\label{table:number_of_works_per_technology}
\end{table}

Table \ref{table:number_of_works_per_technology} presents a summary of the most common technologies, methods, and techniques found in the literature proposing solutions for wearable devices. The second column values refer to the number of papers using the related technology, method, or technique from 2008 to Apr/2017. As with previous categorizations, the same work may be counted in two different categories of problems. Additionally, Table \ref{table:number_of_works_per_technology} shows only the most common categories found in this review. These facts are the main reason why the sum of ``Number of works'' differs from the total number of evaluated papers.

\begin{figure*}[!htbp]
  \centering
  \includegraphics[scale = 0.6]{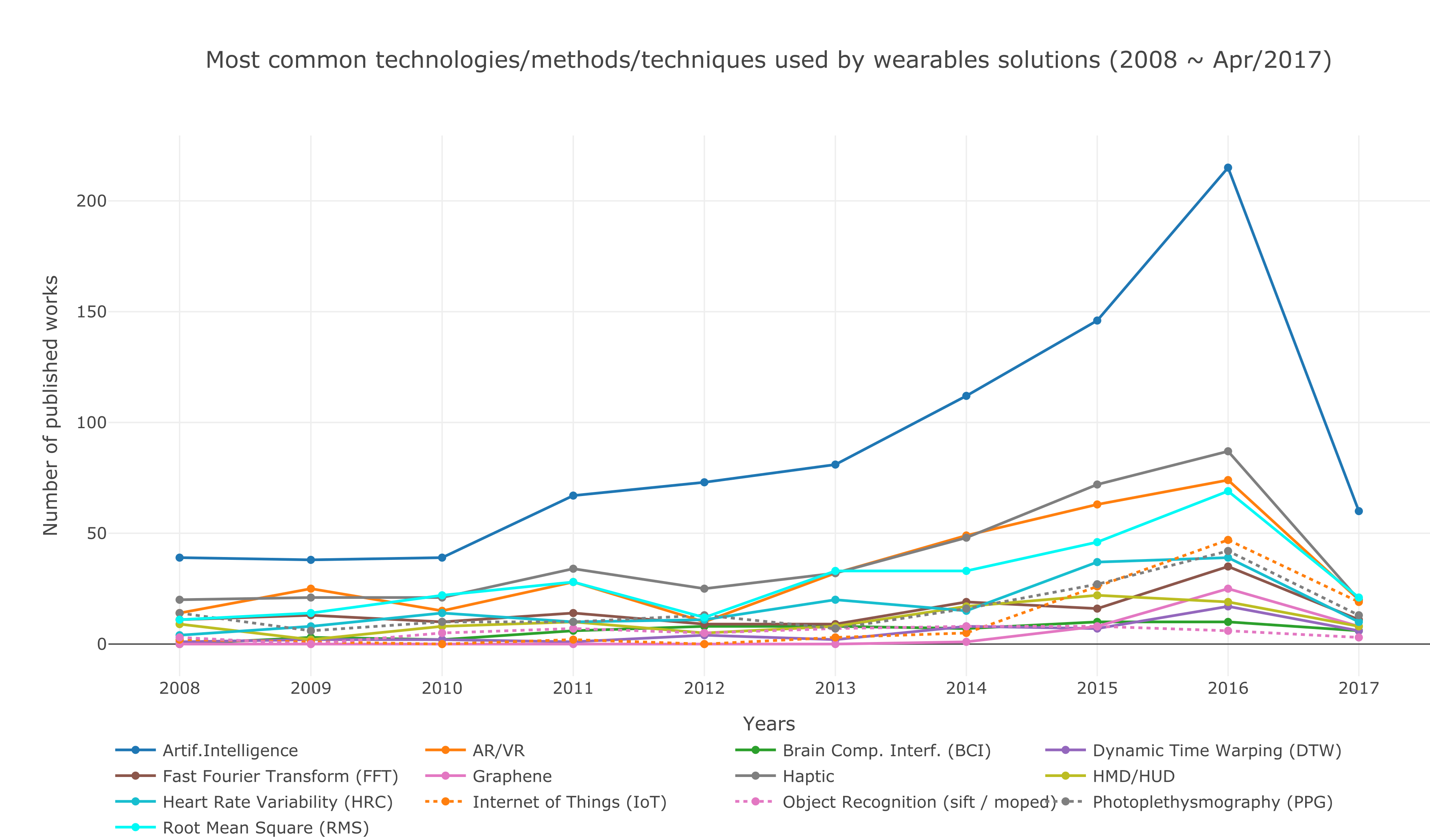}
  \caption{Most common technologies/methods and techniques addressed by wearable research during the considered period.}
  \label{fig:wearables_devices_technologies}
\end{figure*}

Information presented in Table \ref{table:number_of_works_per_technology} shows a great number of proposals making use of ``Artificial Intelligence'' technologies. More than 25\% of all studies refer to this area. This trend has strengthened in recent years owing to the resurgence of artificial intelligence, as depicted in Figure \ref{fig:wearables_devices_technologies} between 2010/2011 and 2015/2016.

Figure \ref{fig:wearables_devices_technologies} also provides relevant insight into how the use of technologies, methods, and techniques applied to the wearable devices development has evolved over the past years. Aside from the ``Artif. Intelligence'' technology, three other technologies can be featured in the chart. The use of ``Haptic'' and ``AR/VR'' technologies in wearable devices has increased over the years, especially during the 2015/2017 period, where their popularity surpassed the remaining items. The same behavior can be correlated to works using ``Root Mean Square (RMS)'', which is a measure commonly applied to signal processing techniques. This information can be proven by analyzing the data presented in \ref{subsubsec:applicability_areas} and \ref{subsubsec:addressed_problems} subsections, highlighting prominent studies on related areas, such as: ``Healthcare/medicine'', ``Environment/individual sensing'' for ``Applicability areas'' and ``Gait assistance/support/tracking'', ``Posture and gesture detection'', and ``Diseases/disturbances detection'' for ``Addressed Problems''.

During the initial years considered by this SRL, the numbers of studies associated with all technologies, methods, and techniques were close to each other. However, this has changed in recent years, where three prominent groups can be seen in the chart: 1) Artificial Intelligence, 2) Haptic devices, AR/VR devices, RMS, and 3) Remaining items. 

It is valid to highlight the use of the ``Heart Rate Variability (HRV)'' measure by wearable devices. Works published during 2008 made few references to the use of such a measure. However, this scenario has changed during the last years. Now, the HRV measure is broadly used in current investigations. One possible reason for this behavior could be related to the increasing number of fitness trackers devices (bands, stripes, and related devices), what may motivate new studies using this type of data.

Owing to the high relevance presented by ``Artificial Intelligence'' in comparison with other items, this SRL a more profound investigation was taken. Through the previously extracted data, this technology was analyzed aiming to discover what are the main components that contribute to its popularity, making it appear in so many studies. Table \ref{table:artif_intelligence_components} depicts the main contributors to the increased use of ``Artif. Intelligence'' technology by studies in the wearable devices context.

\begin{table}[!htbp]
\caption{Artif. Intelligence - Main topics screening considering wearable devices studies (2008 -- Apr/2017)}
\begin{center}
\begin{tabular}{ c | l } 
\hline
\multicolumn{1}{c}{\textbf{Artif. Intelligence Topic}} & \multicolumn{1}{c}{\textbf{Number of works}}  \\ 
\hline
\hline
\rowcolor{Gray}Machine Learning & \makebox[0.5cm]{396} \mybar{3}\\ 
Neural Networks & \makebox[0.5cm]{150} \mybar{1,136363636}\\ 
\rowcolor{Gray}Support Vector Machines (SVM) & \makebox[0.5cm]{132} \mybar{1}\\ 
Hidden Markov Models (HMM) & \makebox[0.5cm]{127} \mybar{0,962121212}\\ 
\rowcolor{Gray}K-nearest Neighbors (KNN) & \makebox[0.5cm]{115} \mybar{0,871212121}\\ 
Decision Tree & \makebox[0.5cm]{107} \mybar{0,810606061}\\ 
\rowcolor{Gray}Principal Comp. Analysis (PCA) & \makebox[0.5cm]{102} \mybar{0,772727273}\\ 
Naive Bayes & \makebox[0.5cm]{69} \mybar{0,522727273}\\ 
\rowcolor{Gray}Random Forest & \makebox[0.5cm]{57} \mybar{0,431818182}\\ 
Logistic Regression & \makebox[0.5cm]{40} \mybar{0,303030303}\\ 
\rowcolor{Gray}Crowdsourcing & \makebox[0.5cm]{25} \mybar{0,189393939}\\ 
Haar Cascade Classifier & \makebox[0.5cm]{23} \mybar{0,174242424}\\ 
\rowcolor{Gray}Dynamic Bayesian Nets. (DBN) & \makebox[0.5cm]{21} \mybar{0,159090909}\\ 
Convolutional Networks & \makebox[0.5cm]{15} \mybar{0,113636364}\\ 
\hline
\end{tabular}
\end{center}
\label{table:artif_intelligence_components}
\end{table}

\begin{figure}[!htbp]
  \centering
  \includegraphics[scale = 0.3]{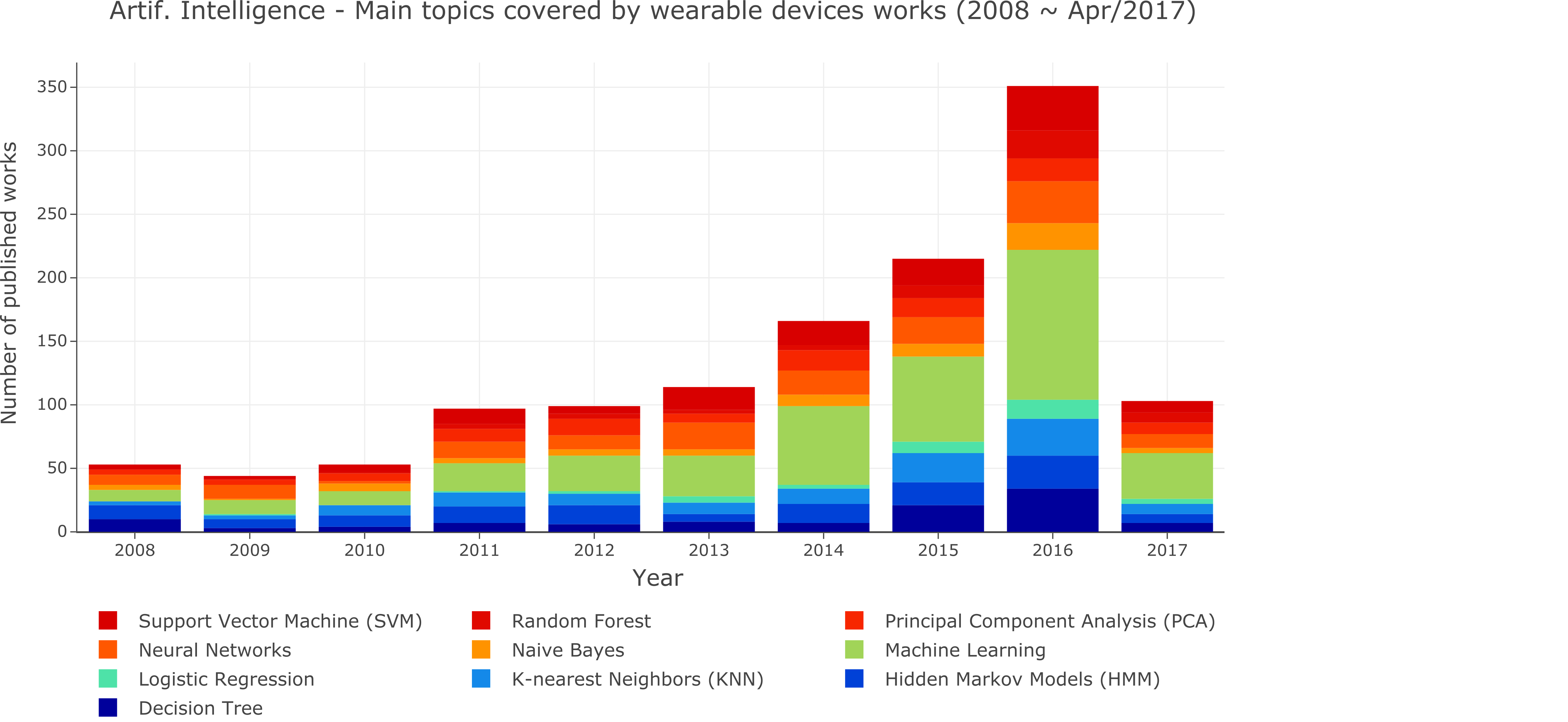}
  \caption{Artif. Intelligence main topics covered by wearable devices researches.}
  \label{fig:artifintelligence_topic}
\end{figure}

Additionally, Figure \ref{fig:artifintelligence_topic} shows a more detailed view of how artificial intelligence use has evolved over the last years in the wearables context. The majority of papers indexed by this SRL make some reference to ``Machine Learning'' techniques/methods, and existing algorithms are presented by both Figure \ref{fig:artifintelligence_topic} and Table \ref{table:artif_intelligence_components}.  

The presented chart also makes it possible to check how artificial intelligence used by wearables evolved from 2008 until 2017. For instance, the use of ``Support Vector Machines'' was scarce in 2008, while the same cannot be said in 2016, with this topic occurring frequently with increasing relevance. The same can be said when analyzing other AI/machine learning approaches, such as: ``K-nearest Neighbors (KNN)'', ``Logistic Regression'', and ``Naive Bayes''.

Finally, the curve shown by the chart is very similar to those presented in ``Applicability areas'' (\ref{subsubsec:applicability_areas}) and ``Addressed problems'' (\ref{subsubsec:addressed_problems}) subsections, suggesting that these research topics have also increased at the same pace as the number of studies. 

\subsubsection{Connectivity \& network} \label{subsubsec:connectivity_res}
Another interesting analysis results from the connectivity information of wearable devices proposed in the studies. The extracted information, mainly the ``Connectivity \& network'', field can be used to classify most common types of networks. This information is particularly interesting for new investigations, as it is possible to verify which types of networks are the most popular for wearables. Additionally, through this data, scientists may choose to conduct their research on less explored types of networks, or even use already developed solutions.

Table \ref{table:wearable_connectivity} shows a categorization of articles reviewed by this SRL according to the network they used. As expected, most devices that make use of a network connection use a ``Bluetooth'' interface. This fact may be explained because this type of network frequently has low energy consumption hardware available, making such interface propitious for wearable device use. 

Moreover, ``Wi-Fi'' and ``Mobile networks'' are presented, respectively, as the second and third most common network interfaces. Currently, these network types are frequently used as hotspots for Internet access or to provide a seamless connection over a large area, despite their constraints in energy consumption, which may impact devices' autonomy.   
      
A note regarding the ``ZigBee'' and ``XBee'' entries should be made. ``ZigBee'', although being specified under the IEEE 802.15 (Bluetooth) standard, was considered here as an independent specification. In the same manner, ``XBee'' is commonly the name given to a set of hardware modules manufactured by the Digi\footnote[1]{https://www.digi.com/xbee} company. However, some solutions presented by the articles make use of these modules as a network interface standard. Thus, to provide a more detailed description of this category, the authors have chosen to present them separately.

\begin{table}[!htbp]
\caption{Connectivity and network types used by wearable devices (2008 -- Apr/2017)}
\begin{center}
\begin{tabular}{ c | l } 
\hline
\multicolumn{1}{c}{\textbf{Network type}} & \multicolumn{1}{c}{\textbf{Number of works}}  \\ 
\hline
\hline
\rowcolor{Gray}Bluetooth & \makebox[0.5cm]{982} \mybar{3}\\ 
Wi-Fi & \makebox[0.5cm]{271} \mybar{0,82790224}\\ 
\rowcolor{Gray}Mobile networks & \makebox[0.5cm]{211} \mybar{0,644602851}\\ 
ZigBee & \makebox[0.5cm]{205} \mybar{0,626272912}\\ 
\rowcolor{Gray}RFID & \makebox[0.5cm]{159} \mybar{0,485743381}\\ 
XBee & \makebox[0.5cm]{63} \mybar{0,192464358}\\ 
\rowcolor{Gray}NFC & \makebox[0.5cm]{32} \mybar{0,097759674}\\ 
Ant+ & \makebox[0.5cm]{2} \mybar{0,00610998}\\ 
\hline
\end{tabular}
\end{center}
\label{table:wearable_connectivity}
\end{table}

\begin{figure*}[!htbp]
 \centering
 \includegraphics[scale = 0.6]{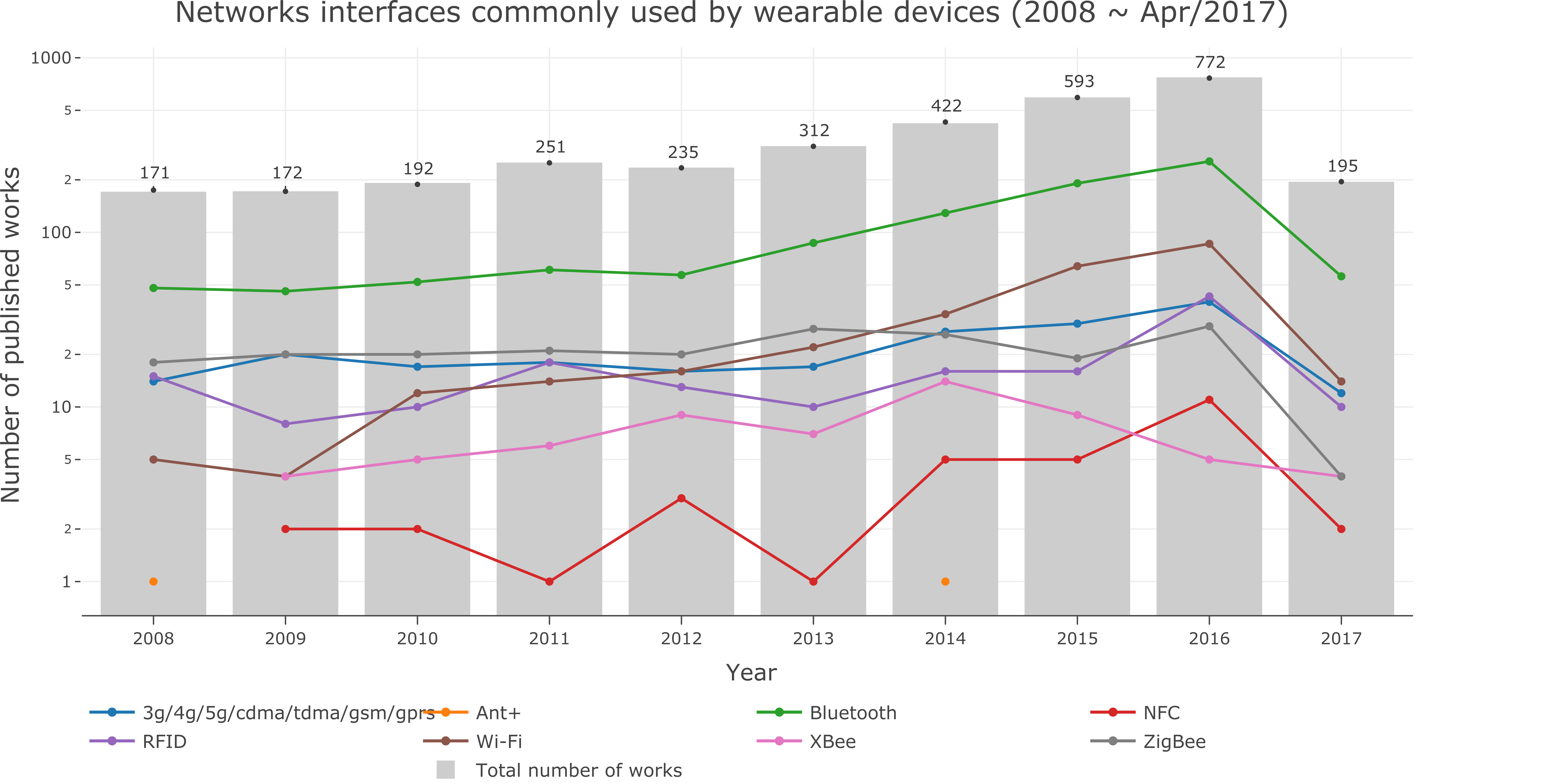}
 \caption{Network types used by wearable devices solutions during the considered period.}
 \label{fig:wearable_connectivity}
\end{figure*}

Figure \ref{fig:wearable_connectivity} presents a discretization, considering recent years, regarding the most used network interfaces. The ``Y'' axis represents the number of solutions using the related interface on a logarithmic basis. The background bars show the total number of works evaluated year-by-year, allowing a comparison metric for the impact of each network interface. Data presented in this chart complement the information previously shown in Table \ref{table:wearable_connectivity}, emphasizing the relevance of ``Bluetooth'' use over the period analyzed. Furthermore, note how the ``Wi-Fi'' network usage has evolved over the past years. In 2008 this network was the fifth most common interface and as of 2017, it takes second place.

Another interesting fact is related to mobile network (``3G/4G/5G/CDMA/TDMA/GSM/GPRS'') popularity. Their use has declined since 2013. However, such networks are now becoming more popular for wearable devices, with their use taking the third position of most popular ones. 

In summary, these results tend to indicate a direct association between energy consumption and mobility levels. While ``Bluetooth'' solutions provide a local network with high mobility level, ``Wi-Fi'' is commonly used as a bridge to deliver wearable device data to the cloud, but imply a low mobility degree. In the chart’s middle and bottom parts, the remaining network types vary from low (``NFC'' and ``ANT+'') to medium (``RFID'' and ``XBee'') usage.

\subsubsection{Hardware components} \label{subsubsec:hardware_components}
During recent years, hardware components have evolved in size through their miniaturization, and hardware cost has also been reduced. Today, a hobbyist developer or scientist has access to high-quality prototyping hardware at a reasonable price, a fact that has helped to increase the number of proposals and solutions in wearable computing. This review has considered a set of hardware components and products used by wearables over the last years, helping to create a specific classification of this subject.

Wearable device prototypes proposed by current studies make use of known hardware components. These components are typically used by hobbyists to create off-the-shelf devices in different applicability areas. As a means to evaluate the most frequently used and trending hardware, this SRL took the extracted data from the ``Hardware components'' field (Table \ref{table:extracted_data}), and categorized the results into different levels to associate them with a problem/applicability area previously described. 

Unlike the prior analyses, hardware data extracted from the articles and presented in Table \ref{table:wearable_hardware} were classified according to the total number of works during the period and the percentage of papers referring to each component. This organization eases the task of determining the popular hardware, and provides a reasonable view of top trending applications. This table indexes a set of highly relevant hardware components retrieved through the ``Data extraction'' step, providing a measure regarding the impact of each component. 

An initial analysis can be made by considering the top-trending items listed in Table \ref{table:wearable_hardware}. The ``Accelerometer sensor'' is listed as being the most referred-to hardware component, being used by 32.46\% of the research prototypes. This information can be better understood looking back to the data presented in subsections \ref{subsubsec:applicability_areas} and \ref{subsubsec:addressed_problems}. There, the ``Healthcare/medicine''/``Environment/individual sensing'' applicability areas and ``Gait assistance/support/tracking''/``Posture and gesture recognition'' problems were listed as the most popular ones. Particularly the second one makes intensive use of Inertial Measurement Units (IMUs) hardware, which consist of accelerometer, gyroscope, and magnetometer sensors. Another interesting point is the relationship of these hardware components with data presented in subsection \ref{subsubsec:device_location_type}, where the most common locations to attach a wearable device are displayed. ``Wrist-worn''/``Arm-worn'' positions can be highlighted as being locations frequently used by devices that sense individual activity (gait, posture, gesture, and so on), consequently making use of IMU sensors.

Moreover, 19.00\% of the works use a ``Camera'' as one of their components. Again, this fact can be justified when considering that many types of devices use this hardware to retrieve images for real-time or post-processing. ``Gait assistance/support/tracking'', ``Gaze tracking'', and ``Eye tracking'' are just some examples of possible application areas where solutions can use a camera. Note the other top-listed hardware components. Almost all can be associated with the ``Healthcare/medicine'' areas, corroborating previously presented results.


\begin{table}[!htbp]
\caption{Most common hardware components used by wearable devices (2008 -- Apr/2017)}
\begin{center}
\begin{tabular}{ c | l | c } 
\hline
\multicolumn{1}{c}{\textbf{Hardware component}} & \multicolumn{1}{c}{\textbf{Number of works}}  & \multicolumn{1}{c}{\textbf{Percentage}}\\ 
\hline
\hline
\rowcolor{Gray}Accelerometer sensor & \makebox[0.5cm]{1076} \mybar{3} & 32.46\%\\ 
Camera & \makebox[0.5cm]{630} \mybar{1,756505576} & 19,00\%\\ 
\rowcolor{Gray}ECG sensor & \makebox[0.5cm]{512} \mybar{1,427509294} & 15.44\%\\ 
Electrodes & \makebox[0.5cm]{507} \mybar{1,413568773} & 15,29\%\\ 
\rowcolor{Gray}Gyro sensor & \makebox[0.5cm]{494} \mybar{1,37732342} & 14.90\%\\ 
Heart rate sensor & \makebox[0.5cm]{437} \mybar{1,218401487} & 13,18\%\\ 
\rowcolor{Gray}Microcontroller & \makebox[0.5cm]{407} \mybar{1,134758364} & 12.28\%\\ 
Impedance analyzer & \makebox[0.5cm]{225} \mybar{0,62732342} & 6,79\%\\ 
\rowcolor{Gray}GPS module & \makebox[0.5cm]{219} \mybar{0,610594796} & 6.61\%\\ 
A/D converter & \makebox[0.5cm]{211} \mybar{0,588289963} & 6,36\%\\ 
\rowcolor{Gray}Pressure sensor & \makebox[0.5cm]{192} \mybar{0,535315985} & 5.79\%\\ 
Arduino & \makebox[0.5cm]{184} \mybar{0,513011152} & 5,55\%\\ 
\rowcolor{Gray}Magnetic sensor & \makebox[0.5cm]{161} \mybar{0,448884758} & 4.86\%\\ 
EEG sensor & \makebox[0.5cm]{149} \mybar{0,415427509} & 4,49\%\\ 
\rowcolor{Gray}Antenna & \makebox[0.5cm]{137} \mybar{0,38197026} & 4,13\%\\ 
Temperature sensor & \makebox[0.5cm]{123} \mybar{0,342936803} & 3,71\%\\ 
\rowcolor{Gray}PPG sensor & \makebox[0.5cm]{122} \mybar{0,340148699} & 3.68\%\\ 
Printed Circuit Board & \makebox[0.5cm]{114} \mybar{0,317843866} & 3,44\%\\ 
\rowcolor{Gray}Microphone & \makebox[0.5cm]{113} \mybar{0,315055762} & 3.41\%\\ 
Piezoelectric comp. & \makebox[0.5cm]{112} \mybar{0,312267658} & 3,38\%\\ 
\rowcolor{Gray}Google Glass & \makebox[0.5cm]{98} \mybar{0,273234201} & 2.96\%\\ 
DC motor & \makebox[0.5cm]{78} \mybar{0,217472119} & 2,35\%\\ 
\rowcolor{Gray}Vibration motor & \makebox[0.5cm]{70} \mybar{0,195167286} & 2.11\%\\ 
Strain sensor & \makebox[0.5cm]{68} \mybar{0,189591078} & 2,05\%\\ 
\rowcolor{Gray}Oximeter & \makebox[0.5cm]{50} \mybar{0,139405204} & 1.51\%\\ 
Potentiometer & \makebox[0.5cm]{49} \mybar{0,1366171} & 1,48\%\\ 
\rowcolor{Gray}Raspberry Pi & \makebox[0.5cm]{33} \mybar{0,092007435} & 0.99\%\\ 
\hline
\end{tabular}
\end{center}
\label{table:wearable_hardware}
\end{table}

\begin{figure*}[!htbp]
 \centering
 \includegraphics[scale = 0.80]{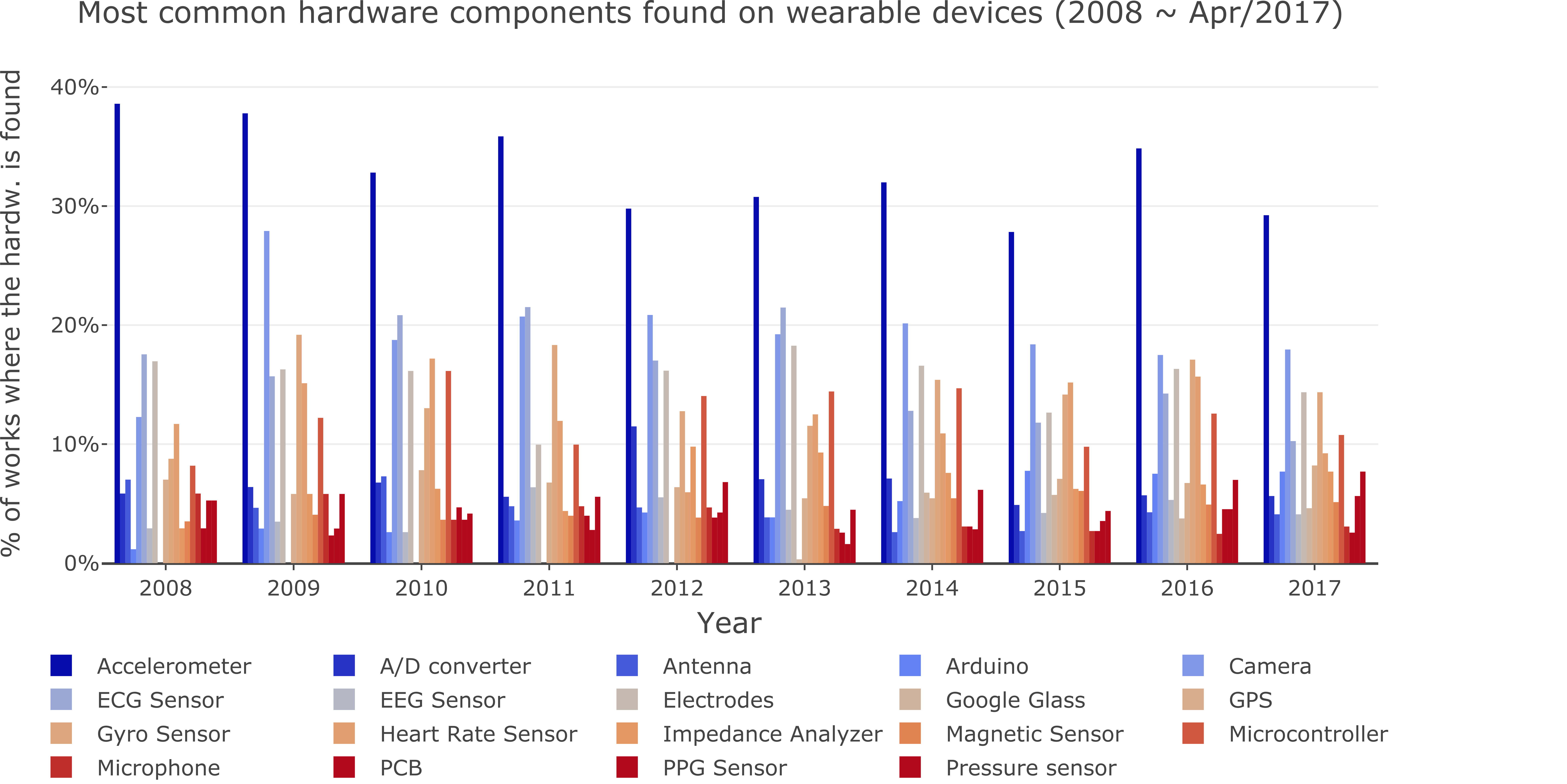}
 \caption{Percentage of wearable devices studies using different hardware components over the considered years.}
 \label{fig:wearable_hardware}
\end{figure*}

Figure \ref{fig:wearable_hardware} shows a chart illustrating the hardware components usage percentage over recent years. As previously presented, ``Accelerometer'' sensor and ``Camera'' were the most referred items. More than 30\% of all articles have made some reference to the former, and almost 20\% of solutions used the latter. This behavior is apparent over all the considered years with minor variations. 

Another impressive result is related to the prototyping platforms. Despite their low relevance, Table \ref{table:wearable_hardware} shows a considerable number of works using ``Arduino'' and ``Raspberry Pi'' boards to create wearable device prototypes. As can be seen in Figure \ref{fig:wearable_hardware}, use of these components has increasingly evolved over the considered years, being employed by hobbyists and scientists within the academic context.

It is also valid to note the advent of the Google Glass device. Until 2014, few articles considered this device when proposing solutions or prototypes. However, after 2014, a more significant number of wearables using Google Glass can be observed. With its reasonable price to end-users, this solution appears to be well accepted by academia as a way to accelerate the development of wearable devices or to validate initial ideas.

\begin{figure}[!htbp]
 \centering
 \includegraphics[scale = 0.26]{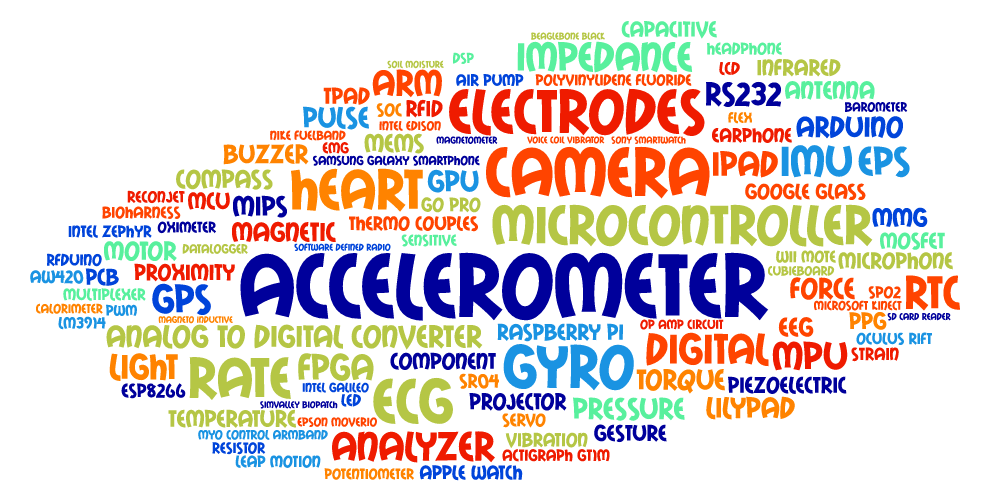}
 \caption{Word cloud presenting wearable devices’ main hardware components relevance.}
 \label{fig:wearable_word_cloud_hardware}
\end{figure}

Although Table \ref{table:wearable_hardware} presents the most frequently used hardware components, this SRL has indexed many other items with low relevance. Figure \ref{fig:wearable_word_cloud_hardware} shows a word cloud of all the indexed hardware, visually showing the impact each element has in comparison with all others. In addition to the information already presented by the referred table, it is also possible to see the most comprehensive set of elements being used, highlighting those top-listed in Table \ref{table:wearable_hardware}. Other components such as ``FPGA'', ``oximeter'', ``ESP8266'', ``compass'', ``buzzer'', and ``Apple watch'' are present, indicating that considerable wider spectrum of components is used by scientists to propose prototypes and solutions. 

A set of final analyses and conclusions can be made through the data presented in this subsection. Third-party solutions are frequently used as a basis to compose more complex prototypes, helping to further expand the wearable computing field. Moreover, the recent popularity of prototyping platforms, such as Raspberry Pi, Arduino, Intel Edison, Intel Galileo, Texas IT Beagle Bone, and so on, have also promoted the design and development of wearables over a new range of solutions.

\subsubsection{Associated layer} \label{subsubsec:associated_layer_res}
\cite{8319168} proposes a novel method to categorize wearable device solutions, associating them to different layers according to their functionalities. This classification starts with the simplest possible solution (Layer 0) and incrementally adds new functionalities until it reaches the most complex ones (Layer 4). It is valid to note that higher layers can reuse features represented by lower layers. A short description regarding the types of devices in each layer is provided. 

\begin{itemize}
    \item \textbf{Layer 0}: Wearables without a visual display. Being the closest to IoT devices, they mainly focus on user sensing, data retrieval, and transmission to a third-party server. Fitness trackers/bands, health monitors, and independent sensors are the most common examples of these devices; 
    \item \textbf{Layer 1}: Devices with a simple visual display that shows 2D images (icons, circles, squares, rectangles and any other pre-processed image), and text. This layer can act to pinpoint a physical place, representing a ``point-of-interest (POI)'';
    \item \textbf{Layer 2}: Wearable solutions with 3D-rendering capabilities that possibly process user interactions in addition to the functionalities provided by lower layers. Graphics enable the device to handle AR and VR applications, providing a set of more realistic applications;
    \item \textbf{Layer 3}: Wearable equipment that allows graphics manipulation on-screen, overlaying real-world images. This layer allows a certain level of interaction with the end-user as a means to increase user experience;
    \item \textbf{Layer 4}: Devices supporting ``Artificial Intelligence (AI)'' techniques, such as machine learning and pattern recognition, to extract, match, and process real-world data/objects. The support of additional hardware may be required to meet real-time constraints.
\end{itemize}


Table \ref{table:wearable_layers_res} describes how wearables can be classified according to the five layers based on the articles considered in this review. Given that ``Layer 0'' is mostly composed of simple devices to sense the environment, it is reasonable to have more than 80\% of all devices included in this set. 

The data presented in Table \ref{table:wearable_layers_res} agree with the results presented in subsections \ref{subsubsec:applicability_areas} and \ref{subsubsec:device_location_type}. The first indicates that the ``Healthcare/medicine'' and ``Environment/individual sensing'' are the most popular applicability areas for wearables, justifying why ``Layer 0'' is the most referred-to, as both areas are composed of devices that retrieve data without the use of a visual display. Moreover, data presented in subsection \ref{subsubsec:device_location_type} help to corroborate these results, as ``Wrist-worn'', ``Arm-worn'', and ``Chest-worn'' are body places frequently used by the solutions, also being the most common positions where a device inside ``Layer 0'' is placed.

The relevant number of solutions classified as ``Layer 3'' can be explained through the same approach. Considering that this layer consists of devices making use of visual interfaces overlaying graphics on the screen, such as glasses and AR/VR goggles, it is easy to understand why it is a popular level. According to the data shown in subsection \ref{subsubsec:common_technologies_methods}, ``AR/VR'' technologies are widely used in wearable solutions, while ``Glasses'' are listed as one of the most popular types of devices in subsection \ref{subsubsec:device_location_type}. 

Layers 1, 2, and 4 are the intermediate points. While ``Layer 1'' holds elements with simple display capabilities, ``Layer 2'' devices cannot be categorized as being complex solutions. In turn, ``Layer 4'' prototypes are more complex and are not as popular as those listed inside ``Layer 3''. It is reasonable to assume that data presented here is simply a snapshot of the current scenario, which may start to change at any time by market or academic demands. 

\begin{table}[!htbp]
\caption{Wearable devices layer distributions (2008 -- Apr/2017)}
\begin{center}
\begin{tabular}{ c | l | c } 
\hline
\multicolumn{1}{c}{\textbf{Layer}} & \multicolumn{1}{c}{\textbf{Number of works}}  & \multicolumn{1}{c}{\textbf{Percentage}}\\ 
\hline
\hline
\rowcolor{Gray}Layer 0 &  \makebox[0.5cm]{2738} \mybarA{0,826}{0,174} & 82.59\%\\ 
Layer 1 & \makebox[0.5cm]{37} \mybarA{0,011}{0,989} & 1,12\%\\ 
\rowcolor{Gray}Layer 2 & \makebox[0.5cm]{116} \mybarA{0,035}{0,965} & 3.50\%\\ 
Layer 3 & \makebox[0.5cm]{359} \mybarA{0,108}{0,892} & 10.83\%\\ 
\rowcolor{Gray}Layer 4 & \makebox[0.5cm]{65} \mybarA{0,020}{0,980} & 1.96\%\\ 
\textbf{Total} & \makebox[0.5cm]{3315} \mybarA{1,00}{0,0} & 100\%\\ 
\hline
\end{tabular}
\end{center}
\label{table:wearable_layers_res}
\end{table} 

\begin{figure}[!htbp]
 \centering
 \includegraphics[scale = 0.28]{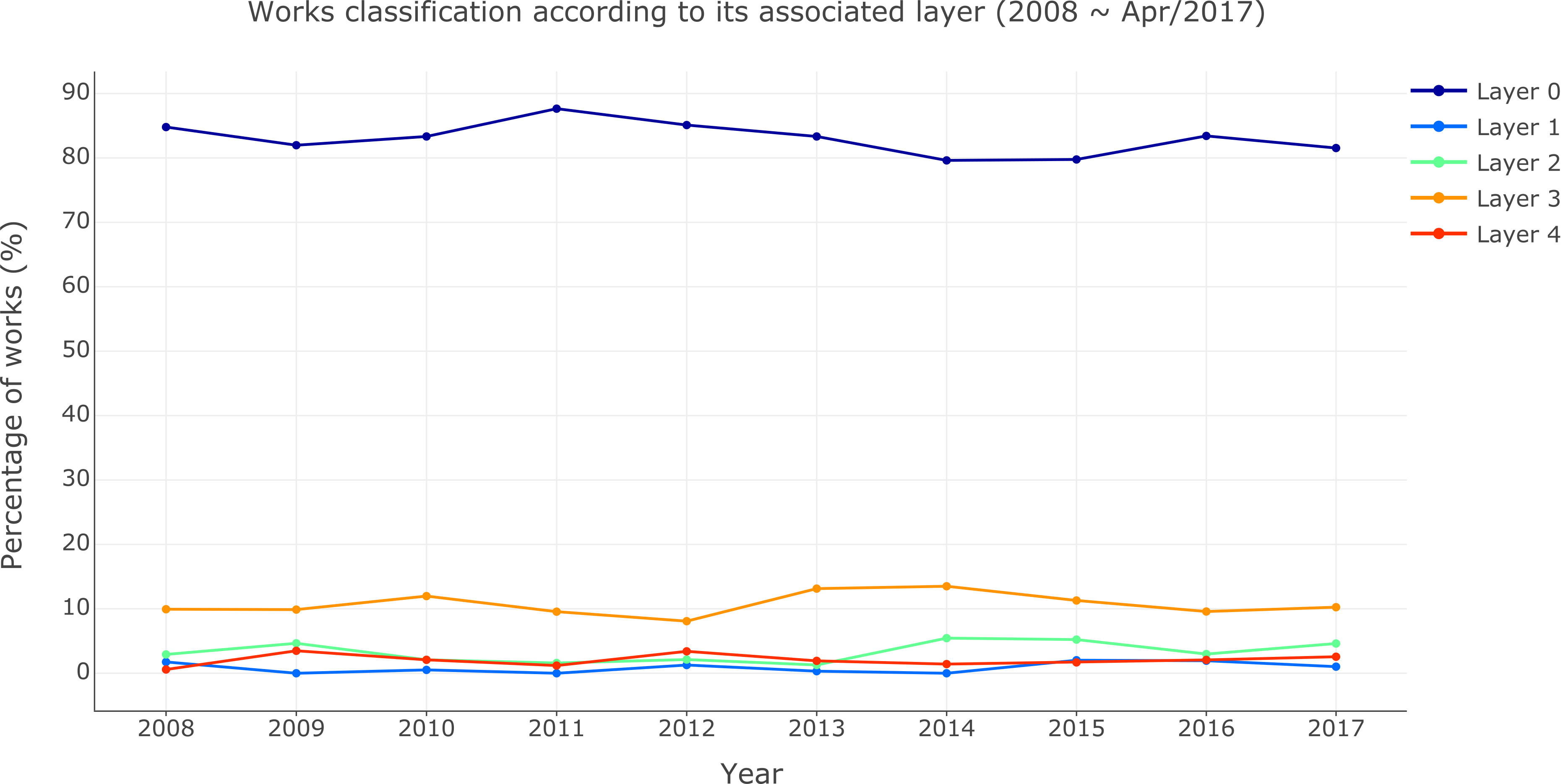}
 \caption{Distribution of wearables solutions according to their layers over the considered years.}
 \label{fig:wearable_layers_res}
\end{figure}

Figure \ref{fig:wearable_layers_res} shows how the wearable devices association with the layers has evolved during the last ten years. The percentage of studies is presented with their related classification for each year. Validating the results presented in Table \ref{table:wearable_layers_res}, the chart shown by this figure also reveals that most of the devices can be associated with ``Layer 0''. The relevant difference here is related to how the number of articles classified as pertaining to each layer has evolved over the past years. A possibly typical case is the comparison of works percentage in ``Layer 2'' versus ``Layer 4'': The number of articles proposing a wearable device in each of these layers was almost the same until 2014. From this time on, the percentage of papers addressing devices classified inside ``Layer 2'' surpassed those in ``Layer 4''. These results can be associated with the advent of virtual reality glasses and their applications during recent years, which moved from scientific experiments to the consumer market, generating an increasing demand for studies in this context.

The chart allows concluding that there is a substantial discrepancy between the number of studies being developed with simple wearable devices in comparison with the most complex solutions. This difference can partially be explained by considering that wearables in ``Layer 0'' most of the times make use of simple sensors and can be applied over a broader range of solutions than any device contextualized into the other layers. Despite this, as the number of studies increases in other areas, such as AR/VR and HUD/HMD, this gap is expected to decrease to some extent.

\section{Concluding Remarks}\label{sec:concluding_remark_and_future_works}
Today, wearable devices on the market provide a considerable number of solutions to be used in several different areas. Solutions applied to entertainment, sports, healthcare, rescue, and other areas can be currently found in stores. In turn, the popularity of these devices challenges scientists to investigate new problems and opportunities, proposing novel solutions and specially raising new questions.

\subsection{Summary}
This systematic review presents a snapshot of current trending topics in the wearables field over the past ten years. The methodology used by this paper allowed the authors to revise and categorize a large set of studies focused on electronic wearable devices. In total, 3,315 papers were found describing different wearable solutions. The data extracted from these papers have allowed a consistent analysis, providing clues about the nature of the work being published and providing relevant information regarding the most trending topics in this area. Based on the number of analyzed papers, the authors believe that the final conclusions faithfully reflect the real, current scenario of this research area.

Through the provided data, it was possible to highlight the increasing interest in wearables technology, with a growth of more than 300\% in the number of published papers since the beginning of the considered period. This initial conclusion leads to a set of new findings, which are presented below. 

\subsection{Main findings}
From the analyses made throughout this review, and considering the scope addressed by each paper, a relevant number of main findings can be listed here.
\begin{itemize}
	\item The majority of research is conducted by universities and research centers in the USA, Japan, and China, respectively. The contribution sum of works provided by these three countries exceeds the sum of all work authored by scientists from the remaining countries;
	\item Despite the number of solutions proposing wearables for environment and individual sensing, healthcare solutions remain the wearable device type’s most targeted applicability area. Thus, corroborating this result, the most relevant problems present a list of topics primarily composed of items related to healthcare and medicine contexts;  
	\item To solve the problems listed by each applicability area, different technologies have been taken into account by the research. This review showed that, since 2010, artificial intelligence has become an important tool with high impact on wearables research. Nevertheless, other relevant technologies were also noted (Haptic, AR/VR, Root Mean Square (RMS), Internet of Things (IoT), ...) reflecting, in some ways, trends raised in other areas;
	\item As a natural consequence of individual sensing, the most common place to attach a wearable is to users' limbs. Although is fairly obvious when analyzing previously presented data, this work has also enumerated the preferred types of wearables and their positions when attached to the human body;
	\item Currently, wearables are highly-connected devices. As expected, the summary presenting the most common network interfaces shows that the majority of connected devices use a ``Bluetooth'' network to send/receive the data; 
	\item Solutions presented by the papers and analyzed by this SRL usually make use of well-known hardware components. Sensors, actuators, prototyping platforms, and other hardware components were listed by this work considering their frequency of occurrence. As a result, a list of most-used items was presented;
	\item Finally, another wearables devices categorization was proposed by this review. Considering a classification that takes complexity and functionality levels into account, proposed devices were separated into five different layers (from the simplest to the most complex ones).
\end{itemize}

This set of findings can be used as a toolbox when designing and building novel solutions. Through the information provided by this SRL, scientists, designers, and engineers have a roadmap describing the most (and the least) explored areas, allowing them to focus their solutions according to the desired result.   

\subsection{Limitations}
Although this review has covered a significant amount of relevant articles, the task to select the proper studies and extract related data takes significant time. Hence, the required work effort to conduct the process, summarize, and finish the review have some influence on the analyzed time window. In the case of this review, almost one year was needed to finalize all the tasks, creating an information gap from the beginning of 2017 until now.

For the reason outlined above, producing an up-to-date SRL is a difficult task, demanding a continuous process of revision. However, a gap is unavoidable because of the execution time needed to conduct the proposed protocol.

\subsection{Future work}
The articles analyzed here provided a set of primary conclusions. However, a deeper screening can be made to find non-obvious and hidden correlations inside the dataset. For instance, the information regarding different countries can be cross-referenced with the types of developed wearables to verify the main interests of each country/research group.

Moreover, considering the already extracted information, a clustering algorithm can be applied to identify the most relevant and actuating research groups, highlighting their main interests and outputs.

Finally, a continuous update can be considered to cover recent works, despite the time required to fulfill this task.

\bibliographystyle{IEEEtran}
\bibliography{bare_conf}
\end{document}